\documentclass{jfm}

\usepackage{graphicx}
\usepackage{newtxtext}
\usepackage{newtxmath}
\usepackage{natbib}
\usepackage{hyperref}
\usepackage{xcolor}
\usepackage{amsmath}

\hypersetup{
    colorlinks = true,
    urlcolor   = blue, %https://www.overleaf.com/project/64834f2e47ddd5e42f2c491f
    citecolor  = black,
}

\newcommand{\RomanNumeralCaps}[1]
\linenumbers

\title{Exact coherent structures in two-dimensional turbulence identified with convolutional autoencoders}
\author{Jacob Page\aff{1},
Joe Holey\aff{2},
Michael P. Brenner\aff{3,4}
and
Rich R. Kerswell\aff{2}}
\affiliation{\aff{1}School of Mathematics, University of Edinburgh, Edinburgh, EH9 3FD, UK, 
\aff{2}DAMTP, Centre for Mathematical Sciences, University of Cambridge, Cambridge, CB3 0WA, UK, 
\aff{3}School of Engineering and Applied Sciences, Harvard University, Cambridge MA 02138, USA,
\aff{4}Google Research, Mountain View, CA 94043, USA.}

\begin{document}
\maketitle

\begin{abstract}
Convolutional autoencoders are used to deconstruct the changing dynamics of two-dimensional Kolmogorov flow as $Re$ is increased from weakly chaotic flow at $Re=40$ to a chaotic state dominated by a domain-filling vortex pair at $Re=400$.
The highly accurate embeddings allow us to visualise the evolving structure of state space and are interpretable using `latent Fourier analysis’ (Page {\em et. al.}, \emph{Phys. Rev. Fluids} \textbf{6}, 2021). 
Individual latent Fourier modes decode into vortical structures with a streamwise lengthscale controlled by the latent wavenumber, $l$, with only a small number $l \lesssim 8$ required to accurately represent the flow. Latent Fourier projections reveal a detached class of bursting events at $Re=40$ which merge with the low-dissipation dynamics as $Re$ is increased to $100$. We use doubly- ($l=2$) or triply- ($l=3$) periodic latent Fourier modes to generate guesses for UPOs (unstable periodic orbits) associated with high-dissipation events. While the doubly-periodic UPOs are representative of the high-dissipation dynamics at $Re=40$, the same class of UPOs move away from the attractor at $Re=100$ — where the associated bursting events typically involve larger-scale ($l=1$) structure too. At $Re=400$ an entirely different embedding structure is formed within the network in which no distinct representations of small-scale vortices are observed; instead the network embeds all snapshots based around a large-scale template for the condensate. We use latent Fourier projections to find an associated `large-scale’ UPO which we believe to be a finite-$Re$ continuation of a solution to the Euler equations.
\end{abstract}

%----------------
%
%  INTRODUCTION
%
%----------------

\section{Introduction}
% Paragraph 1: Low order modelling of turbulence, challenges of dynamically relevant bases etc. Combine with dynamical systems ideas. Big statement sentence etc etc

The dynamical systems view of turbulence \citep{Hopf1948, Eckhardt2002, Kerswell2005,Eckhardt2007,Gibson2008, Cvitanovic2010,Kawahara2012,Suri2020,Graham2021, Crowley2022} has revolutionised our understanding of transitional and weakly turbulent shear flows. 
In this perspective, a realisation of a turbulent flow is considered as a trajectory in a very high-dimensional dynamical system, in which unstable periodic orbits (UPOs) and their stable and unstable manifolds serve as a skeleton for the chaotic dynamics \citep{Hopf1948,ChaosBook}. 
However, progress with these ideas in multiscale turbulence at high Reynolds numbers ($Re$) has been slower, which can be largely attributed to the challenge of finding suitable starting guesses for UPOs to input in a Newton-Raphson solver \citep{Kawahara2001, Viswanath2007, Cvitanovic2010, Chandler2013}.
As such, it is unknown whether a reduced representation of a high-$Re$ flow in terms of UPOs is possible, and how rapidly the number of such solutions grows as $Re$ increases. 
In this work we outline a methodology based on learned embeddings in deep convolutional autoencoders that can both (i) map out the structure of the state space of solutions at a given $Re$ and (ii) generate effective guesses for UPOs that describe the high dissipation bursting dynamics. 
% excellent 

Since the first discovery of a UPO in a (transiently) turbulent Couette flow by \citet{Kawahara2001} there has been a flurry of interest and the convergence of many more UPOs in the same configuration \citep{Viswanath2007,Cvitanovic2010,Page2020} and other simple geometries \citep{Chandler2013, Willis2013}.
Individual UPOs isolate a closed cycle of dynamical events which are also observed transiently in the full turbulence \citep{Wang2007,Hall2010}, although all solutions found to date have been dominated by a single lengthscale, either as domain-filling vortices and streaks \citep{Kawahara2001}, spatially isolated structures \citep{GibsonBrand2014} or attached eddies in unbounded shear \citep{Doohan2019}.
This is in contrast to turbulence at high $Re$ which is inherently multiscale \citep[e.g.][]{Jimenez2012}. 
% - Nice turbulent physics understanding 
% - Relevant to high Re? 

% PO search -- space/time tapestry but fundamental diffculties 
The most popular method for searching for periodic orbits, termed `recurrent flow analysis', relies on a turbulent orbit shadowing a UPO for at least a full cycle, with the near recurrence --measured with an $L_2$ norm-- identifying a guess for both the velocity field and period of the solution \citep{Viswanath2007,Cvitanovic2010,Chandler2013}. 
This inherently restricts the approach to lower $Re$, as the shadowing becomes increasingly unlikely as the Reynolds number is increased due to the increased instability of the UPOs. 
Furthermore, measuring near recurrence with an Euclidean norm is unlikely to be a suitable choice for a distance metric on the solution manifold unless the near recurrence is very close \citep{Page2021}. 
More recent methods have sought to remove near recurrence, for example by using dynamic mode decomposition \citep{Schmid2010} to identify the signature of nearby periodic solutions \citep{Page2020,Marensi2022}, or by using variational methods that start with a closed loop as an initial guess \citep{Lan2004,Parker2022}. 

% Low order PO ideas 
% Paragraph : Standard low-order modelling of turbulence vs more modern approaches -- PCA, DMD vs more complex AEs (refs to Brunton/Koumoutsakos/standard AE stuff). 
Periodic orbit theory has rigorously established how the statistics of UPOs can be combined to make statistical predictions for chaotic attractors in strictly hyperbolic systems \citep{Artuso1990a,Artuso1990b}. 
The hope in spatiotemporal chaos (i.e. turbulence) is that a similar approach may be effective even with an incomplete set of UPOs.
For instance, see the statistical reconstructions in \citet{PNBK2022} or the Markovian models of weak turbulence in \citet{Yalnuz2021}. 
This approach is attractive because it allows one to unambiguously `weight' individual dynamical processes in their contribution to the long-time statistics of the flow.
However, the ability to apply these ideas at high-$Re$ is currently limited by the search methods described above, while extrapolating results at low $Re$ upwards is challenging because of the emergence of new solutions in saddle-node bifurcations, or the turning-back of solution branches \citep[e.g.][]{Gibson2008,Chandler2013}. 
% The hope with searching for UPOs has been that they 
% Statistical basis of periodic orbits -- CK13, PNBK22arXiv -- and low order models (Burak Markvov model).
% But come back to outstanding question of finding solutions; 
% paper/KF a natural candidate to explore complexification of state space under increasing Re, because of (i) computational simplicity, (ii) known solutions, (iii) connection with Euler in high Re limit. 

The challenge of extending the dynamical systems approach to high-$Re$ also involves a computational element associated with both the increased grid requirements and the slowdown of the `GMRES' aspect of the Newton algorithm used to perform an approximate inverse of the Jacobian \citep{Veen2019}. 
Modern data-driven and machine learning techniques can play a role here, for instance by reducing underlying resolution requirements through learned derivative stencils \citep{Kochkov2021}, or by building effective low-order models of the flow \citep[e.g. see the overview in][]{Brunton2020}. 
In dynamical systems, neural networks have been highly effective in estimating attractor dimensions \citep{Linot2020}, and more recently in building low order models of weakly turbulent flows \citep{JesusGraham2023,Linot2023}.
One aspect of the latter study \citep{Linot2023} that is particularly promising is that the low-order model (a `neural' system of differential equations) was used to find periodic orbits that corresponded to true UPOs of the Navier-Stokes equation.
However, all of these studies have only been performed at modest $Re$ or in other simpler PDEs.
Our aim here is to build a low-order model that can be applied over a wide range of $Re$.

In previous work \citep[][hereafter PBK21]{Page2021}, we used a deep convolutional neural network architecture to examine the state space of a two-dimensional turbulent Kolmogorov flow (monochromatically forced on the 2-torus) in a weakly turbulent regime at $Re=40$. 
By exploiting a continuous symmetry in the flow we performed a `latent Fourier analysis' of the embeddings of vorticity fields, demonstrating that the network built a representation around a set of spatially-periodic patterns which strongly resembled known (and unknown) unstable equilibria and travelling wave solutions. 
In this paper we consider the same flow but use a more advanced architecture to construct low order models over a wide range of $40 \leq Re \leq 400$. 
Latent Fourier analysis provides great insight into the changing structure of the state space as $Re$ increases, by, for example, showing how high-dissipation bursts merge with the low dissipation dynamics beyond the weakly chaotic flow at $Re=40$. 
It also allows us to isolate hundreds of new UPOs associated with high dissipation events that have not been found by any previous approach. 

Two-dimensional turbulence is a computationally attractive testing ground for these ideas because of the reduced computational requirements and the ability to apply state-of-the-art neural network architectures which have been built with image classification in mind \citep{LeCun2015,huang2017densely}. 
Kolmogorov flow itself has been widely studied with very large numbers of simple invariant solutions documented in the literature \citep{Chandler2013,Lucas2014,Lucas2015,Farazmand2016,Parker2022}, although the phenomenology in 2D is distinct from three-dimensional wall bounded flows discussed above due to the absence of a dissipative anomaly and the inverse cascade of energy to large scales \citep{Onsager1949,Smith1993,Kraichnan1980,Boffetta2012}.
There is the intriguing possibility that UPOs at high-$Re$ in this flow actually connect to (unforced) solutions of the Euler equation \citep{Zhigunov2023}, and we are able to use our embeddings to explore this effect here by designing a UPO-search strategy for structures with particular streamwise scales. 
Our analysis highlights the role of small-scale dynamical events in high-dissipation dynamics, and also suggests that the UPOs needed to describe turbulence at high-$Re$ may need to be combined in space as well as time, a viewpoint which is consistent with idea of a `spatio-temporal' tiling advocated by \citet{Gudorf2019}.
% There are still open questions in this flow about the physical mechanisms of the inverse cascade, although in Kolmogorov flow is typically forced at fairly large scale which does not allow for a significant inertial range. 

% General 2D phenomenology \citep{Onsager1949, Smith1993, Kraichnan1980, Boffetta2012} -- inverse cascade, large scale vortices, presumed link to Euler equations in simple invariant solutions \citep{Zhigunov2022}. 
The remainder of the manuscript is structured as follows: 
In \S\ref{sec:config} we describe the flow configuration and datasets at the various $Re$, along with a summary of vortex statistics under increasing $Re$. We also outline a new architecture and training procedure that can accurately represent high dissipation events in all the flows considered. 
In \S\ref{sec:lfa} we perform a latent Fourier analysis for three values of $Re$, generating low-dimensional visualisations of the state space to examine the changing role of large scale patterns and the emergence of the condensate.
\S \ref{sec:upos} summarises our UPO search, where we perform a modified recurrent flow analysis at $Re=40$ before using the latent Fourier modes themselves to find large numbers of new high dissipation UPOs.
Finally, conclusions are provided in \S\ref{sec:conc}.

%
% Fig 1
%
\section{Flow configuration and neural networks}
\label{sec:config}
\subsection{Kolmogorov flow}
\begin{figure}
    \centering
    \includegraphics[width=\textwidth]{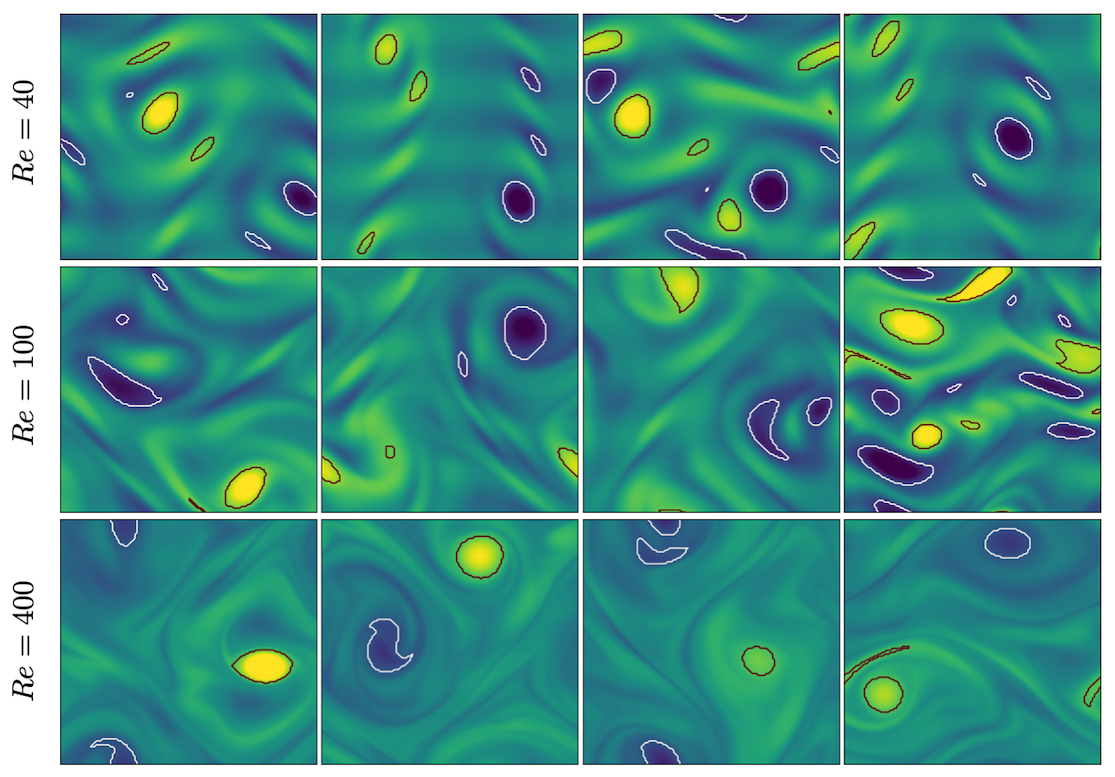}
    \includegraphics[width=\textwidth]{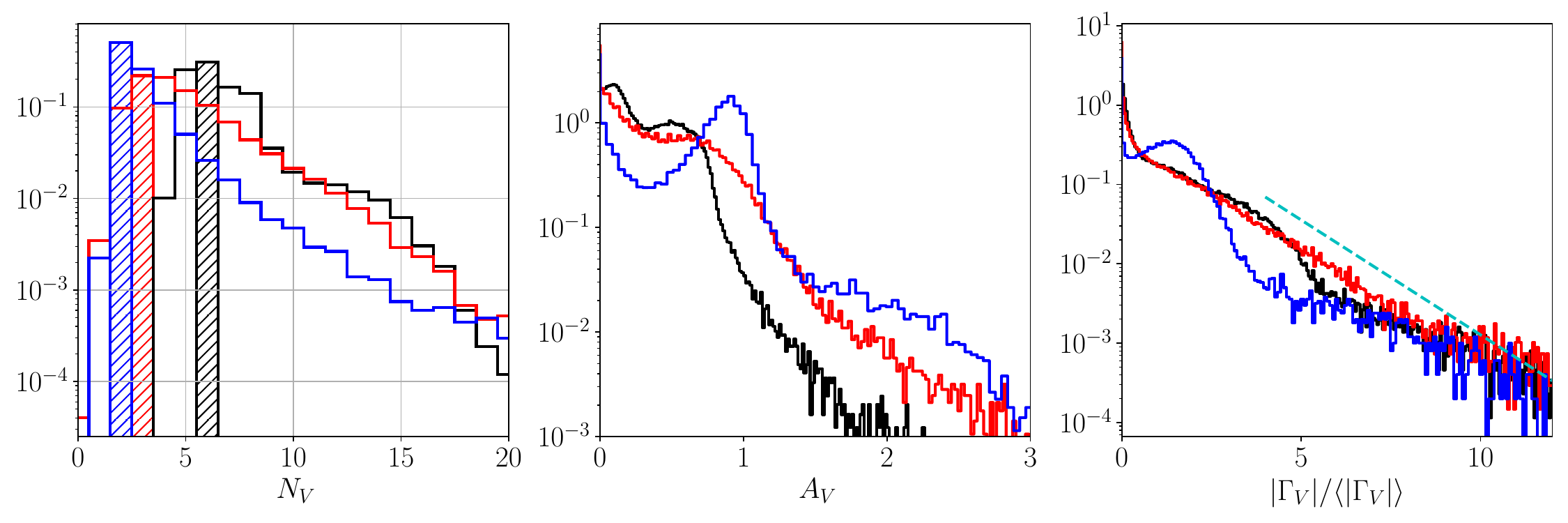}
    \caption{
    % 4 snapshots at each $Re$ and a dissipation PDF for each (excluding 80).
    % Contour levels from top to bottom, $\pm 8$, $\pm 10$, $\pm 20$.
    % Vortex statistics for $Re\in \{40, 100, 400\}$ ($Re=80$ largely indistinguishable from $Re=100$.
    Vorticity snapshots and statistics for $Re=40$, $100$ and $400$. Results at $Re=80$ are qualitatively very similar to those at $Re=100$ and are not shown. 
    Top three rows show snapshots from within the test dataset at $Re=40$ (top; max contour levels $\pm 8$), $Re=100$ (centre; max contour levels $\pm 10$) and $Re=400$ (bottom; max contour levels $\pm 20$). Black/white lines indicate connected regions (`vortices' discussed in the text) where $|\omega(\mathbf x, t) - \langle \omega \rangle| \geq 2\omega_{\text{RMS}}$.
    Bottom row summarises the vortex statistics in the test datasets at $Re=40$ (black), $100$ (red) and $400$ (blue), from left to right showing p.d.fs of numbers of vortices (modal contribution highlighted with the vertical bars), vortex area and normalised vortex circulation (dashed line is $\exp\left(-(2/3) |\Gamma_V| / \langle |\Gamma_V|\rangle\right)$). 
    }
    \label{fig:kf_overview}
\end{figure}
% Training and test datasets; data augmentation. 
We consider two-dimensional flow on the surface of a 2-torus, driven by a monochromatic body force in the streamwise direction (`Kolmogorov' flow). 
The out-of-plane vorticity satisfies
\begin{equation}
    \partial_t \omega + \boldsymbol u \cdot \boldsymbol \nabla \omega = \frac{1}{Re} \Delta \omega - n\cos \, ny,
    \label{eqn:vorticity_kf}
\end{equation}
where we have used the amplitude of the forcing, $\chi$, and the fundamental vertical wavenumber of the box, $k= 2\pi / L_y$, to define a lengthscale, $k^{-1}$, and timescale, $\sqrt{1/(k \chi)}$, so that $Re:=\sqrt{\chi}k^{-3/2}/\nu $.
Throughout we set $L_x = L_y$ and the forcing wavenumber $n=4$ as in previous work \citep[][PBK21]{Chandler2013}. 

Equation (\ref{eqn:vorticity_kf}) is equivariant under continuous shifts in the streamwise direction, $\mathscr T_s: \omega(x,y) \to \omega(x+s,y)$, under shift-reflects by a half wavelength in $y$, $\mathscr S: \omega(x, y) \to -\omega(-x, y+\pi/4)$ and under a rotation by $\pi$, $\mathscr R: \omega(x,y) \to \omega(-x,-y)$. 
In contrast with other recent studies \citep{Linot2020,JesusGraham2023} we explicitly do not perform symmetry reduction. The retention of the continuous symmetry is central to our methods for interpreting the learned embeddings (discussed below). 

We consider a range of Reynolds numbers, $Re\in \{40, 80, 100, 400\}$. 
Our training and test datasets are generated with the open-source, fully differentiable flow solver \texttt{jax-cfd} \citep{Kochkov2021}. 
The workflow for converging UPOs evolved during the course of the project: we primarily converge periodic orbits in the spectral version of \texttt{jax-cfd} \citep{Dresdner2022}, though some results at $Re=40$ were obtained with an in-house spectral code \citep{Chandler2013,Lucas2014}. 
%Discussion of the numerics is included in appendix REF. 
Resolution requirements were adjusted based on the specific $Re$ considered, ranging from $256^2$ at $Re=40$ (in the finite difference version of \texttt{jax-cfd}) to $1024^2$ at $Re=400$. 
Lower resolutions (by a factor of 2) were used in the spectral solvers when converging UPOs.
We downsampled our higher $Re$ training and test data to a resolution of $N_x \times N_y = 128 \times 128$ for consistent input to the neural networks. 
% Some example snapshots at each value of $Re$ are reported in figure \ref{fig:kf_overview} alongside probability density functions of the dissipation. 

% \emph{\color{blue}{}Please verify...}
At each $Re \in \{40, 80, 100\}$ we generated a training dataset by initialising $1000$ trajectories from random initial conditions, discarding an initial transient before saving $100$ snapshots from each, with snapshots separated by an advective time unit. 
When training the neural networks (discussed below) we applied a random symmetry transform to each of the $N = 10^5$ snapshots ($\omega \to \mathscr T^{\alpha}\mathscr S^m \mathscr R^q \omega$, with $\alpha \in [0, 2\pi)$, $m \in \{0, 1,\dots,7\}$ and $q \in \{0,1\}$) each time we looped through the dataset.
We also generated test datasets of the same size in the same manner. 
For the highest $Re=400$, our training dataset was smaller and formed of $700$ trajectories and our test dataset at $Re=400$ consisted of $200$ trajectories. 
In training (network architecture discussed below) we reserved $10\%$ of the training data for validation to avoid overfitting.

The range of $Re$ considered ranges from a weakly chaotic flow at $Re=40$ to the formation of a pair of large scale vortices that dominate the flow --the `condensate' \citep{Onsager1949,Smith1993}-- at $Re=400$. 
These qualitative differences in the flow are explored in figure \ref{fig:kf_overview} where we report snapshots at $Re\in \{40, 100, 400\}$ ($Re=80$ is largely indistinguishable from $Re=100$) along with some statistical analysis of the vortical structures present in the flow. 
The vortex statistics shown here are computed by first computing the root-mean-square vorticity fluctuations $\omega_{\text{RMS}} := \sqrt{\langle \left(\omega(\mathbf x, t) - \overline{\omega} (y)\right)^2\rangle}$, where $\langle \bullet \rangle$ represents a time-, spatial and ensemble-average over the trajectories in the test dataset, while $\overline{\bullet}$ is a time-, $x-$ (`horizontal') and ensemble-average. 
We then extract spatially localised `vortices' as connected regions where $|\omega(\mathbf x, t) - \langle \omega \rangle| \geq 2\omega_{\text{RMS}}$.

The vortex statistics shown in figure \ref{fig:kf_overview} are consistent with the emergence of the condensate at $Re=400$: at this point the flow spends 50\% of its time in a state where there are a pair of vortices, and often higher numbers of vortices, $N_V > 2$, actually indicates a state like that shown in the third and fourth snapshots at $Re=400$ where a small-scale region of high shear qualifies as a `vortex’ as described above. 
These observations are clearly supported by a peak in both the vortex area and circulation probability distribution functions at $Re=400$. 
In contrast, the statistics at $Re=40$ and $Re=100$ do not indicate the dominance of a single large-scale coherent state, but instead the vortex statistics are qualitatively similar to those reported in the early stages of decaying two-dimensional turbulence reported by \citet{Jimenez2020} -- see dashed line in the circulation statistics shown in figure \ref{fig:kf_overview}. 

As observed in earlier studies \citep[][PBK21]{Chandler2013}, the `turbulence' at $Re=40$ is only weakly chaotic and spends much of its time in a state which is qualitatively similar to the first non-trivial structure to bifurcate off the laminar solution at $Re \approx 10$, but with intermittent occurrences of more complex high-dissipation structures. 
In contrast, the dynamics at $Re=100$ are much richer and display an interplay between larger-scale structures and small-scale dynamics.
The increasing dominance of large-scale structure as $Re$ increases can be tied to the emergence of new simple invariant solutions, which can presumably be connected to solutions of the Euler equations as $Re\to \infty$ \citep{Zhigunov2023}. 
Similarly, the decreasing role of smaller scale vortical events in this limit can be associated with the movement of a set of small-scale UPOs away from the attractor. 
Our aim is to use learned embeddings within deep autoencoders to explore this process, mapping out the structure of the state space under increasing $Re$ and finding the associated UPOs. 
% We will now construct robust low-order representations of these dynamics to 
% We will now show how well-designed neural networks can be used to extract UPOs that NICE DESCRIPTION and allow us to assess the range of dynamical events that operate in the flow under increasing $Re$ prior to the emergence of the condensate.

\subsection{DenseNet autoencoders}
\label{sec:densenet}
We construct low-dimensional representations of Kolmogorov flow by training a family of deep convolutional autoencoders, $\{\mathscr A_m^{Re}\}$, which seek to reconstruct their inputs
\begin{equation}
    \mathscr A_m^{Re} (\omega) \equiv [\mathscr D_m^{Re} \circ \mathscr E_m^{Re}](\omega) \approx \omega.
\end{equation}
The autoencoder takes an input (scalar) vorticity field (essentially a greyscale image) and constructs a low-dimensional embedding via and encoder function, $\mathscr E_m^{Re}: \mathbb R^{N_x\times N_y} \to \mathbb R^m$, before a decoder converts the embedding back into a vorticity snapshot, $\mathscr D_m^{Re}: \mathbb R^m \to \mathbb R^{N_x\times N_y}$.

The architecture trained in PBK21 performed well at $Re=40$, but we were unable to obtain satisfactory performance at higher $Re$ with the same network. 
% \rk{[The following sentence needs a rewrite - I'm not sure of the meaning]} {\em Indeed, even at $Re=40$, the reconstruction of high dissipation snapshots it notably weakened relative to the simpler quiescent dynamical events when using the older model.} 
Even the relatively strong performance at $Re=40$ in the PBK21 model was skewed towards low-dissipation snapshots, with the performance on the high dissipation events being substantially weaker. 
To address this performance issue, we: 1. designed a new architecture with a more complex graph structure and feature map shapes motivated by discrete symmetries in the system; and 2. trained the network with a modified loss function to encourage a good representation of the rarer, high dissipation events. 

The structure of the new autoencoder is a purely convolutional network, with dimensionality reduction performed via max pooling as in PBK21. However, we use so-called ``dense blocks’’ \citep{huang2017densely,huang2019convolutional} in place of single convolutional layers, allowing for increasingly abstract features as the outputs of multiple previous convolutions are concatenated prior to the next convolution operation. 
Our dense blocks (described fully in appendix \ref{sec:app_network}) are each made up of three individual convolutions, with the output feature maps of each convolution then concatenated with the input. 
This means that the feature maps at a given scale (before a pooling operation is applied) can be much richer than a single convolution operation. 
For instance, if the input to a particular dense block is an image with $K$ channels, and each convolution adds 32 features, then the output of the block is an image with $K + 3 \times 32$ channels. 
For comparison, a standard convolution with 32 filters of size $N_x’ \times N_y’$ on the same image would be specified by $O(32 \times N_x’ \times N_y’ \times K)$ parameters, while the dense block described here requires $O(32 \times N_x’ \times N_y’ \times (96 + 3K))$ parameters due to the repeated concatenation with the upstream input feature maps.
We also made other minor modifications to the network that are detailed in appendix \ref{sec:app_network}. 
At the inner-most level, the network represents the input snapshot with a set of $M$ feature maps of shape $4 \times 8$, where the `8’ (corresponding to the physical y direction) is fixed by the 8-fold shift-reflect symmetry, $\mathscr S^8 \omega \equiv \omega$, in the system.
The restriction to purely convolutional layers and the smallest feature map size constrains the dimension of the latent space to be a multiple of 32. 
Overall our new model is roughly twice as complex as that outlined in PBK21, with $\sim 2.15 \times 10^6$ trainable parameters for the largest models ($m=1024$) -- the increased cost of the dense blocks being offset somewhat by the absence of any fully connected layers. 
For context, training for 500 epochs on $10^5$ vorticity snapshots (see appendix \ref{sec:app_network}) takes roughly 48 hours on a single NVIDIA A100 GPU (80$\,$GB memory). 

We train the networks to minimise the following loss:
\begin{align}
    \mathscr L := \frac{\gamma}{N}\sum_j \| \mathscr A_m (\omega_j) - \omega_j \|^2 + \frac{(1-\gamma)}{N}\sum_j \| \mathscr A_m^2 (\omega_j) - \omega_j^2 \|^2,
    \label{eqn:full_loss}
\end{align}
over $500$ epochs \citep[batch size of $64$, constant learning rate in an Adam optimizer $\eta=5 \times 10^{-4}$][]{Kingma2015}.
% An additional term is included with the standard `mean squared error’ to force the network to learn a reasonable representation of rarer, high dissipation events. The hope is that the low dimensional representations generated will allow the nature of bursting dynamics to be probed as $Re$ is increased.
An additional term has been added to the standard `mean squared error’ in the loss function (\ref{eqn:full_loss}). 
The new term is essentially a mean squared error on the square of the vorticity field — high dissipation events are associated with large values of the enstrophy $\iint \omega^2 d^2 \mathbf x$, and the new term in (\ref{eqn:full_loss}) makes the strongest contribution to the overall loss in these cases, while quiescent, low dissipation snapshots are dominated by the standard mean squared error term.  
The rationale here is to encourage the network to learn a reasonable representation of high dissipation events, particularly as we increase $Re$. 
Treatment via a modified loss is required as these events make up only a small fraction of the training dataset — a similar effect could perhaps be anticipated if the training data was drawn from a modified distribution skewed to high dissipation events rather than sampled from the invariant measure. 
Intriguingly, we found that the addition of this term lead to better overall mean squared error than a network trained to minimise the mean squared error alone. % verify this is true on the new architecture! 
We trained independent networks at $Re=40$ with various $\gamma$ and found $\gamma=1/2$ to be most effective. We fix $\gamma$ to this value for all networks and $Re$. 

%
% Fig 2
%
\begin{figure}
    \centering
    \includegraphics[width=0.6\textwidth]{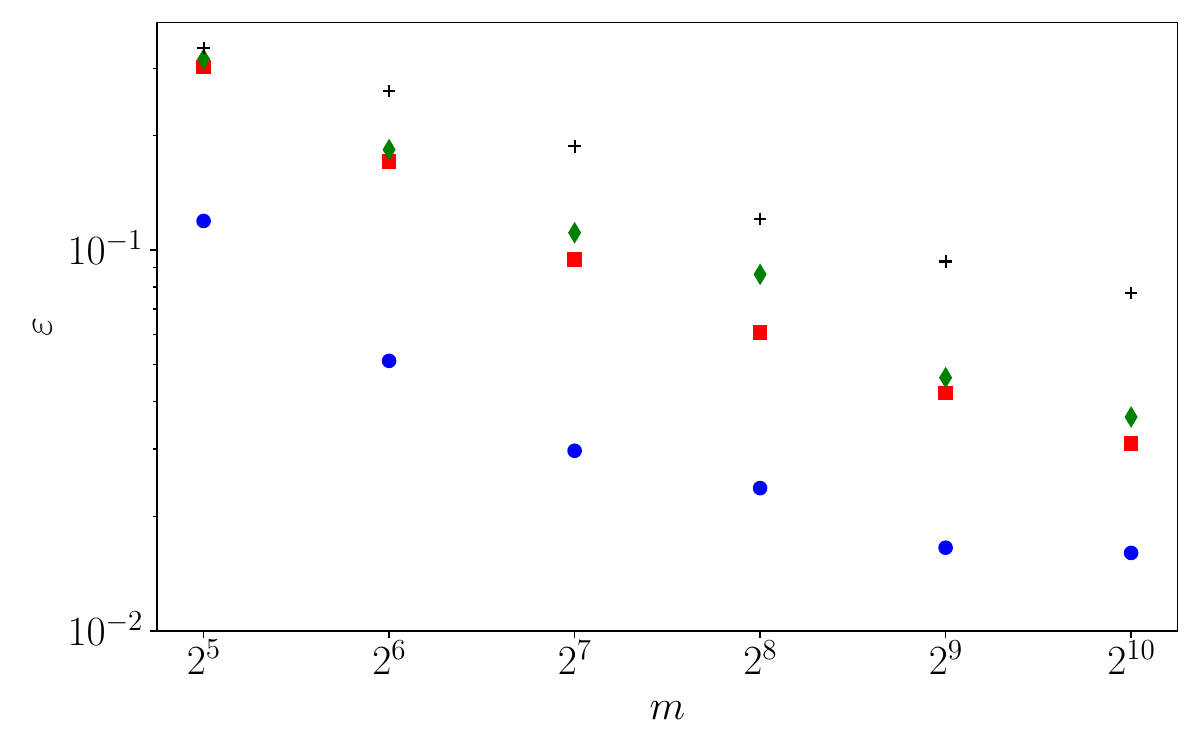}
    \caption{Average relative error (\ref{eqn:local_err}) over the test set(s) as a function of embedding dimension for the best examples of all models trained in this work (recall the latent space dimension has a multiple of $32$ by design). Blue circles for $Re=40$; red squares for $Re=80$; green diamonds for $Re=100$ and black crosses for $Re=400$.}
    \label{fig:ae_overview}
\end{figure}
\begin{figure}
    \centering
    \includegraphics[width=\textwidth]{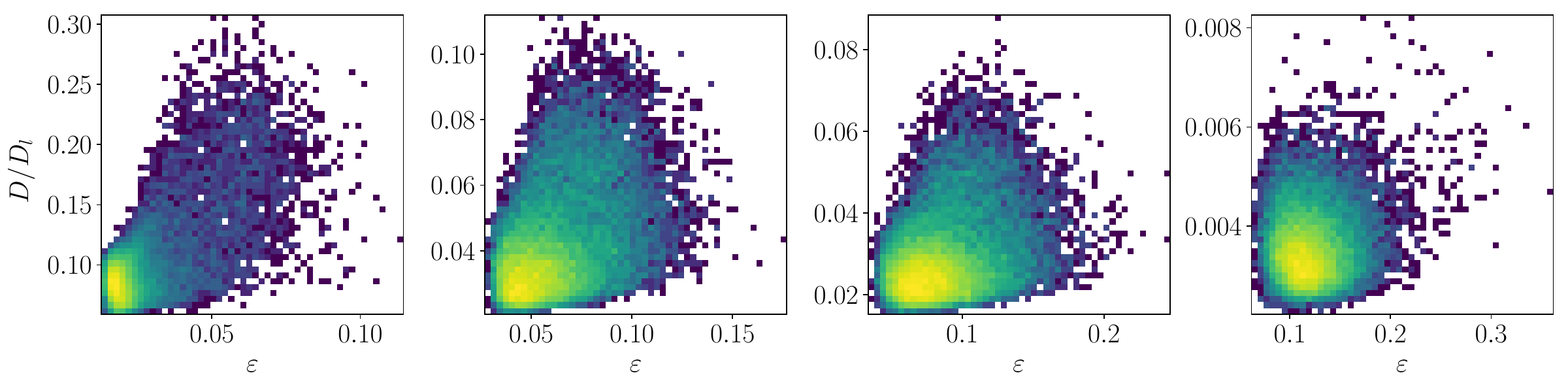}
    \caption{
    Histograms of the test datasets for all $Re \in \{40, 80, 100, 400\}$ ($Re$ increasing left to right in the figure), visualised in terms of the relative error (\ref{eqn:local_err}) computed using the $m=256$ networks and the snapshot dissipation value normalised by the laminar value. 
    The form of the loss function used in training (\ref{eqn:full_loss}) ensures that the rarer high dissipation events are embedded to a similar standard to the low dissipation data which makes up a much larger proportion of the observations.
    }
    \label{fig:loss_diss_pdfs}
\end{figure}

The networks were trained by normalising the inputs $\omega \to \omega / \omega_{\text{norm}}$, where $\omega_{\text{norm}} = 25$ for $Re \in \{40, 80, 100\}$ and $\omega_{\text{norm}} = 60$ for $Re=400$. 
For a consistent analysis of the performance across the architectures and $Re$, we do not report the loss (\ref{eqn:full_loss}) directly, but instead compute the relative error for each snapshot
\begin{equation}
	\varepsilon_j := \frac{\| \omega_j - \mathscr A_m^{Re}(\omega_j) \|}{\| \omega_j \|}
 \label{eqn:local_err}
\end{equation}
where the norm $\| \omega \| := \sqrt{(1/4\pi^2) \iint \omega^2 d^2 \mathbf x }$. 
The average error over the test set for our new architecture is reported for a range of embedding dimensions $m \in \{32, 64, \dots, 1024\}$ and all $Re$ in figure \ref{fig:ae_overview}. 
Unsurprisingly, there is monotonic reduction in the error with increasing network dimensionality $m$ at fixed $Re$, and monotonic increase in the error with increasing $Re$ at fixed $m$.
The best performing network at $Re=40$ shows an average relative error of $\sim 1$\% (exact test-set mean error $0.016$, and standard deviation $0.0076$), while the performance at $Re=400$ has dropped to $\sim 8$\% (mean test-set error of $0.077$ and standard deviation $0.017$). 
% \rk{[Does this mean $\varepsilon_j \approx 0.01 \, \forall j$? should we be more precise e.g. what happens to the average and standard deviation over all snapshots?]}
Visually, snapshots from all networks are very hard to distinguish from the ground truth once $\varepsilon_j \lesssim 0.1$. %\rk{[how do I interpret $\varepsilon$ with no subcript? a representative value? see last line in  this section too]} % JP yes just representative here but will be more precise in the discussion above
For comparison, the same error measure evaluated on the induced velocity fields computed from $\omega$ and $\mathscr A_m(\omega)$ is $\sim 2\%$ for snapshots where the vorticity error is $\varepsilon_j \approx 10\%$ at $Re=400$, and can be as much as a full order of magnitude better. 

The distribution of errors over the test dataset is explored in more detail for the $m=256$ networks at all four Reynolds numbers in figure \ref{fig:loss_diss_pdfs}, where we examine the dependence of reconstruction error on the dissipation of the snapshot. 
Notably, there is no significant loss of performance on the rarer, high dissipation events. 
This is to be contrasted with the sequential autoencoders considered in PBK21 and was achieved using the dual loss function (\ref{eqn:full_loss}) which encourages a robust embedding of snapshots with stronger enstrophy. 
We exploit the quality of the high-dissipation embeddings here to explore the nature of high dissipation events as $Re$ increases.

For the remainder of this paper we consider three networks, $(Re, m) = (40, 128)$, $(100, 512)$ and $(400, 1024)$. 
This covers the full range of $Re$ for which we have trained models (excluding $Re=80$ which is qualitatively very similiar to $Re=100$), with embedding dimensions $\{m\}$ selected to balance model performance against interpretability. 
Each of the three networks yields average relative errors $\varepsilon$ of between roughly $2-8\%$. 

%
% 3
%
\section{Dynamical processes and recurrent patterns}
\label{sec:lfa}

%
% 3.1
%
\subsection{Latent Fourier analysis} 

PBK21 introduced `latent Fourier analysis' as a method to interpret the latent representations within neural networks for systems exhibiting a continuous symmetry.
We will use the same approach here to understand how the state space of Kolmogorov flow complexifies under increasing $Re$, with a particular focus on the high-dissipation ``bursting'' events.
We briefly outline the numerical procedure for performing a latent Fourier decomposition of an encoded vorticity field, $\mathscr E_m(\omega)$.

To perform a latent Fourier analysis we seek a linear operator that can perform a continuous streamwise shift in the latent space
\begin{equation}
    \mathbf T_{\alpha} \mathscr E_m (\omega) := \mathscr E_m (\mathscr T_{\alpha} \omega),
    \label{eqn:shift_op}
\end{equation}
where $\alpha$ is a chosen shift in the $x$-direction. 
An approximate latent-shift operator $\widehat{\mathbf T}_{\alpha}$ is found via a least-squares minimisation \citep[i.e. dynamic mode decomposition:][]{Rowley2009,Schmid2010} over the test set of embedding vectors and their $x$-shifted counterparts:
\begin{equation}
    \widehat{\mathbf T}_{\alpha} = \mathbf E_m(\{\mathscr T_{\alpha} \omega\}) \mathbf E_m(\{\omega\})^+ \, \in {\mathbb R}^{m \times m},
    %\begin{bmatrix}\mathscr E_m(\mathscr T_{\alpha} \omega_1) & \cdots & \mathscr E_m(\mathscr T_{\alpha} \omega_N) \end{bmatrix}
    %\begin{bmatrix}\mathscr E_m(\omega_1) & \cdots & \mathscr E_m(\omega_N) \end{bmatrix}^+,
\end{equation}
where the $+$ indicates a Moore-Penrose psuedo-inverse and 
$$\mathbf E_m(\{\omega\}) := \begin{bmatrix}\mathscr E_m(\omega_1) & \cdots & \mathscr E_m(\omega_N) \, 
\end{bmatrix} \in {\mathbb R}^{m \times N} $$
with $N \gg m$ ($N = O(10^5)$). % [give typical value for $N$]?.
As in PBK21, we consider a shift $\alpha = 2\pi / n$, with $n\in \mathbb N$, and given that $\mathbf T^n_{\alpha}\mathscr E_m(\omega) = \mathscr E_m(\omega)$ due to streamwise periodicity, the eigenvalues of the discrete latent-shift operator are $\Lambda_j = \exp(2\pi i l_j/n)$, where $l_j\in \mathbb Z$ is a \emph{latent} wavenumber.
The parameter $n$ is incrementally increased (i.e. reducing $\alpha$) until we stop recovering new latent wavenumbers. 
Across our networks, we find that only a handful of latent wavenumbers are required -- substantially less than the number of wavenumbers required to accurately resolve the flow.
For example, $l_{\text{max}}=3$ for the $m=128$ network at $Re=40$, while the much higher dimensional network $m=1024$ at $Re=400$ uses wavenumbers as high as $l_{\text{max}}=8$. 

With the non-zero latent wavenumbers determined we can now write down an expression for an embedding of a snapshot subject to an arbitrary shift, $s\in \mathbb R$, 
%\rk{[sure you want $s$ instead of already introduced $\alpha$?]} % JP: Yes, because alpha is fixed and defines the discrete operator, s is continuous 
in the streamwise direction:
\begin{equation}
    \mathscr E_m(\mathscr T_s \omega) = \sum_{l = -l_{\text{max}}}^{l_{\text{max}}} \left(\sum_{k=1}^{d(l)} \mathscr P^l_k(\mathscr E_m(\omega))\right)e^{ils}
    \label{eqn:latent_ft}
\end{equation}
where
\begin{equation}
\mathscr P^l_k(\mathscr E_m(\omega)):=
\biggl[(\boldsymbol \xi_k^{(l)\dagger})^H \mathscr E_m (\omega) \biggr]\boldsymbol \xi^{(l)}_k
\end{equation}
with  $\boldsymbol \xi^{(l)}_k$ the $k^{th}$ eigenvalue in the subspace $l$ (and $\boldsymbol \xi^{(l)\dagger}_k$ is the corresponding adjoint eigenfunction so $(\boldsymbol \xi_i^{(l)\dagger})^H\boldsymbol \xi^{(l)}_j = \delta_{ij}$),
which makes the connection with a Fourier transform clear.
%
% \rk{[btw defn in PBK21 which uses $j=0,...,d(l)-1$] rather than $k=1,...d(l)$ also slightly different symbol for projection - p.s. I see we went to `s' there...]}
% JP -- yes, would prefer to keep new notation (not sure why I wanted to index from 0 previously, must have been thinking in code)
%
The small number of non-zero latent wavenumbers means that each must encode a wide variety of different patterns in the flow -- 
each latent wavenumber is (potentially highly) degenerate with geometric multiplicity $d(l)$. % -- compare the small number required (e.g. $l_{\text{max}}=3$ at $Re=40$) to the large number of streamwise wavenumbers in physical space.
In equation (\ref{eqn:latent_ft}) the quantity $\mathscr P^l_k(\mathscr E_m(\omega))$ is the projection of the embedding vector onto direction $k$ within the eigenspace of latent wavenumber $l$. We obtain these projectors via a SVD within a given eigenspace, which is described  below in \S\ref{3.2}.
We refer to the decode of a projection onto individual latent wavenumbers as a `recurrent pattern'.

At the innermost representation in the autoencoders the turbulent flow is represented with a set of feature maps of shape $4 \times 8$, where the convolutional operations mean that the 4 horizontal cells each correspond to a quarter of the original domain -- i.e. the network has constructed some highly abstract feature that was originally located in one quarter of the physical domain in $x$. 
Similarly, the 8 vertical cells isolate some feature located in one of eight vertical bands in the original image. 
The choice of a particular feature map size for the encoder can be used to encourage the network to learn recurrent patterns at a particular scale. 
This architectural choice is at the root of the relatively low values of $l_{\text{max}}$ observed for the networks. 
For example, consider the case of a single feature map for the encoder -- this would correspond to the $m=32$ networks trained here. 
In this case, we are effectively coarse graining the original vorticity snapshot to a single $4\times 8$ image. 
Therefore, by the Nyquist criterion the maximum wavenumber that can be resolved in the horizontal direction is $l_{\text{max}} = 2$: the network has to learn to couple smaller scale features to large scales in the most efficient way. 
By expanding the number of feature maps at the inner most level, as is done for the $m > 32$ networks (note that the $m=1024$ network has 32 feature maps), we in principle allow for higher latent wavenumbers, but it is seemingly inefficient for the network to embed features in this way, and instead most of the energy is contained in relatively low $l$.
This is a benefit from an interpretability point of view, since individual recurrent patterns have a physical significance; each features a large number of physical wavenumbers with a base periodicity set by the value of $l$.

%
% Fig 4
%
\begin{figure}
    \centering
    \includegraphics[width=0.7\textwidth]{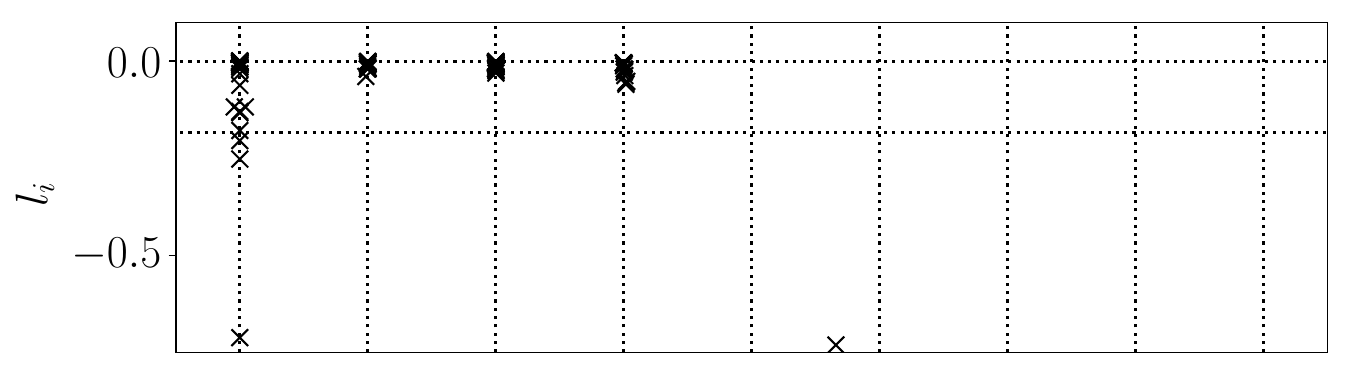}
    \includegraphics[width=0.7\textwidth]{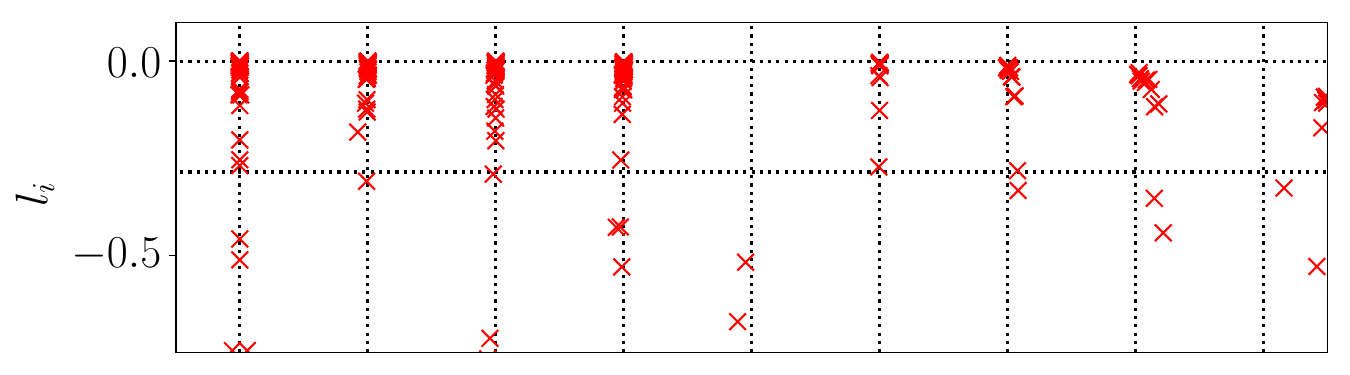}
    \includegraphics[width=0.7\textwidth]{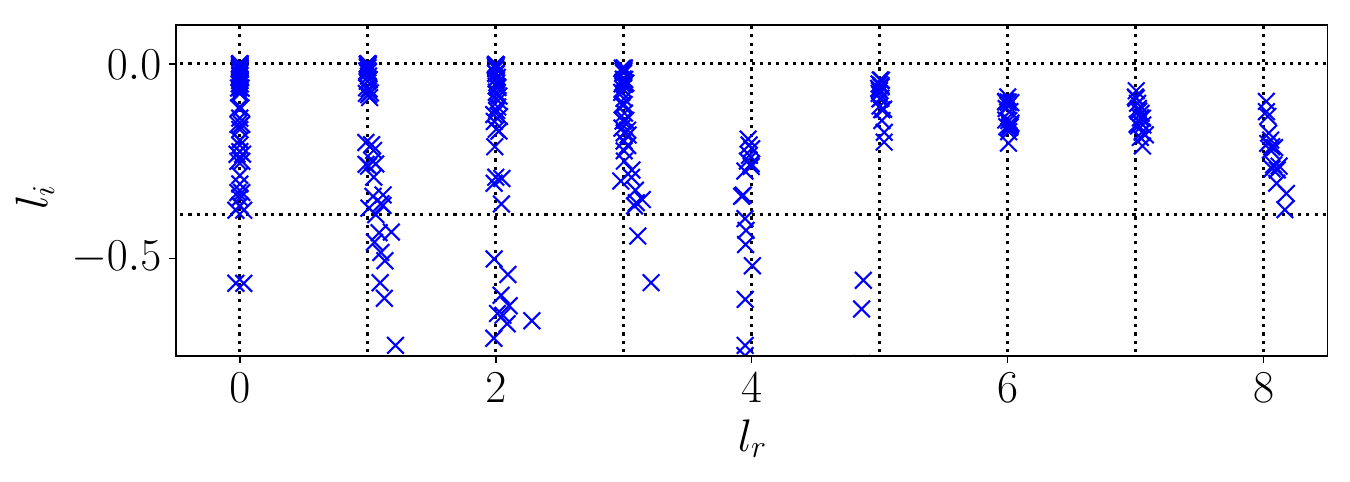}
    \caption{Eigenvalues of the latent shift operator, $\widehat{\mathbf T}_{\alpha}$, visualised in terms of latent wavenumbers $l = l_r + i l_i$ (in a perfect approximation $l_i=0$). 
    (Top) Network $(Re,m) = (40, 128)$, with shift $\alpha = 2\pi / 11$;
    (middle) Network $(Re,m) = (100, 512)$, with shift $\alpha = 2\pi / 17$;
    (bottom Network $(Re, m) = (400, 1024)$, with shift $\alpha = 2\pi / 23$. 
    Latent wavenumbers are determined by writing the eigenvalues of the shift operator as $\Lambda_j = \exp(2\pi i l_j/n)$, where $\alpha \equiv 2\pi / n$. 
    Horizontal dotted lines indicate the threshold $|\Lambda| = 0.9$, or equivalently $l_i = -n\log(0.9)/2\pi$, below which eigenvalues are discarded when doing latent projections. Dotted vertical lines indicate integer $l$.}
    \label{fig:latent_wavenumbers}
\end{figure}

We report eigenvalue spectra for the latent wavenumbers in figure \ref{fig:latent_wavenumbers} for each of the three networks considered here. 
The clustering of points around the integer values of $l$ shows that each wavenumber is highly degenerate.
The maximum latent wavenumber found (and the size of the eigenspaces) expands with increasing $Re$, and consequently a smaller value of $\alpha$ is needed to construct the shift operator (\ref{eqn:shift_op}).
There are large numbers of points clustered on integer values of $l$ in figure \ref{fig:latent_wavenumbers}, though we also find that some structure in the network can not be accurately shifted with a linear operator, and we also observe `decaying' eigenvalues with negative imaginary part. 
This behaviour is more apparent as $Re$ increases. 

%
% Fig 5
%
\begin{figure}
    \centering
    \includegraphics[width=0.344\textwidth]{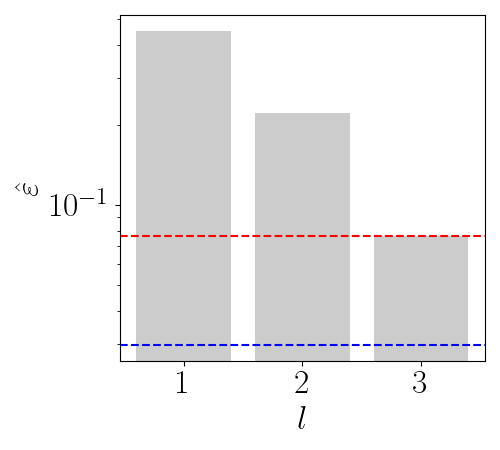}
    \includegraphics[width=0.31\textwidth]{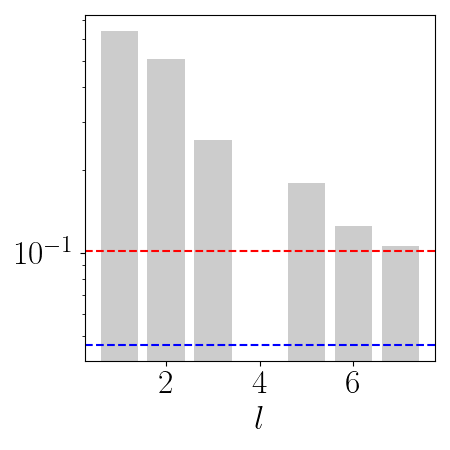}
    \includegraphics[width=0.31\textwidth]{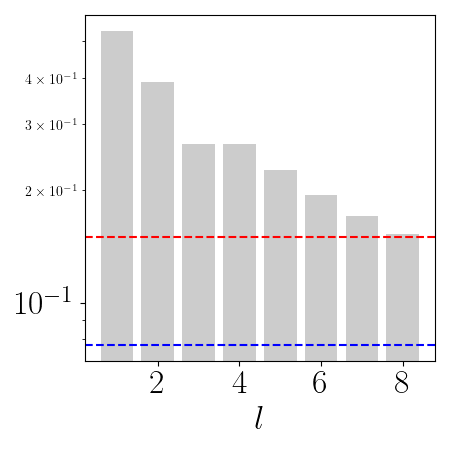}
    \caption{
    Test-set average error in reconstruction (equation \ref{eqn:recon_err_l}) as a function of increasing $l$.
    Dashed blue line is the error using the full latent representation as reported in figure \ref{fig:ae_overview}, dashed red line is the error when reconstruction is performed using all eigenvectors of $\widehat{\mathbf T}_{\alpha}$ with eigenvalues $|\Lambda| > 0.9$. 
    From left to right: $(Re,m) = (40, 128)$, $(100, 512)$ and  $(400, 1024)$. 
    Note that there are no patterns within $l=4$ subspace at $Re=100$ (see figure \ref{fig:latent_wavenumbers}). 
    %\rk{[what's happened to $l=4$ in the middle plot - really missing?! btw abscissa labelling not so helpful as we are only dealing in integers.]}
    }
    \label{fig:wavenumber_completeness}
\end{figure}

Despite the expanding $l_{\text{max}}$ with increasing $Re$ and the increasing inaccuracy of the shift operator, all networks can produce a fairly robust representation of the flow with just three non-zero $l$. 
The reconstruction accuracy of a truncated set of recurrent patterns is examined in figure \ref{fig:wavenumber_completeness}, where we report the average reconstruction error as the number of latent wavenumbers used is incrementally increased.
This error is defined per snapshot as
\begin{equation}
    \hat{\varepsilon}_j(l') := \frac{\| \omega_j - [\mathscr D \circ \widehat{\mathscr{E}}^{l'}](\omega_j)\|}{\|\omega_j\|},
    \label{eqn:recon_err_l}
\end{equation}
where 
\begin{equation}
	\widehat{\mathscr E}^{l'}(\omega) := \sum_{l = -l'}^{l'} \left(\sum_{k=1}^{d(l)} \mathscr P^l_k(\mathscr E(\omega))\right),
\end{equation}
is the projection onto the first $l' > 0$ latent wavenumbers.
Note that we retain only eigendirections for which the associated eigenvalue $|\Lambda_j| > 0.9$. 

The error reported in figure \ref{fig:wavenumber_completeness} (equation \ref{eqn:recon_err_l} averaged over the test set) monochromatically drops as the maximum latent wavenumber is increased -- though note the loss in respresentation from discarding wavenumbers with eigenvalues $|\Lambda| < 0.9$. 
% While the $m=512$ network as does have latent wavenumbers as high as $l_{max} =7$ (see figure \ref{fig:wavenumber_completeness}), only the first Y are required to reduce the reconstruction error to X\% of that obtained with the full encoding retained. 
For the flow configuration considered here, it is really the first three non-zero latent subspaces that do most of the `heavy lifting', and they encode a huge variety of dynamical processes. 
The error with $l' = 3$ at $Re=400$ is $O(0.2)$, which still produces snapshots which are visually hard to distinguish from the input (see earlier discussion of the error in \S\ref{sec:densenet}). 
% We will explore this further below where we use the latent projections to identify small scale periodic orbits which are representative of isolated spatial structures observed in bursts. 

%
% 3.2
%
\subsection{Recurrent patterns}
\label{3.2}
%
% Fig 6
%
\begin{figure}
    \centering
    \includegraphics[width=\textwidth]{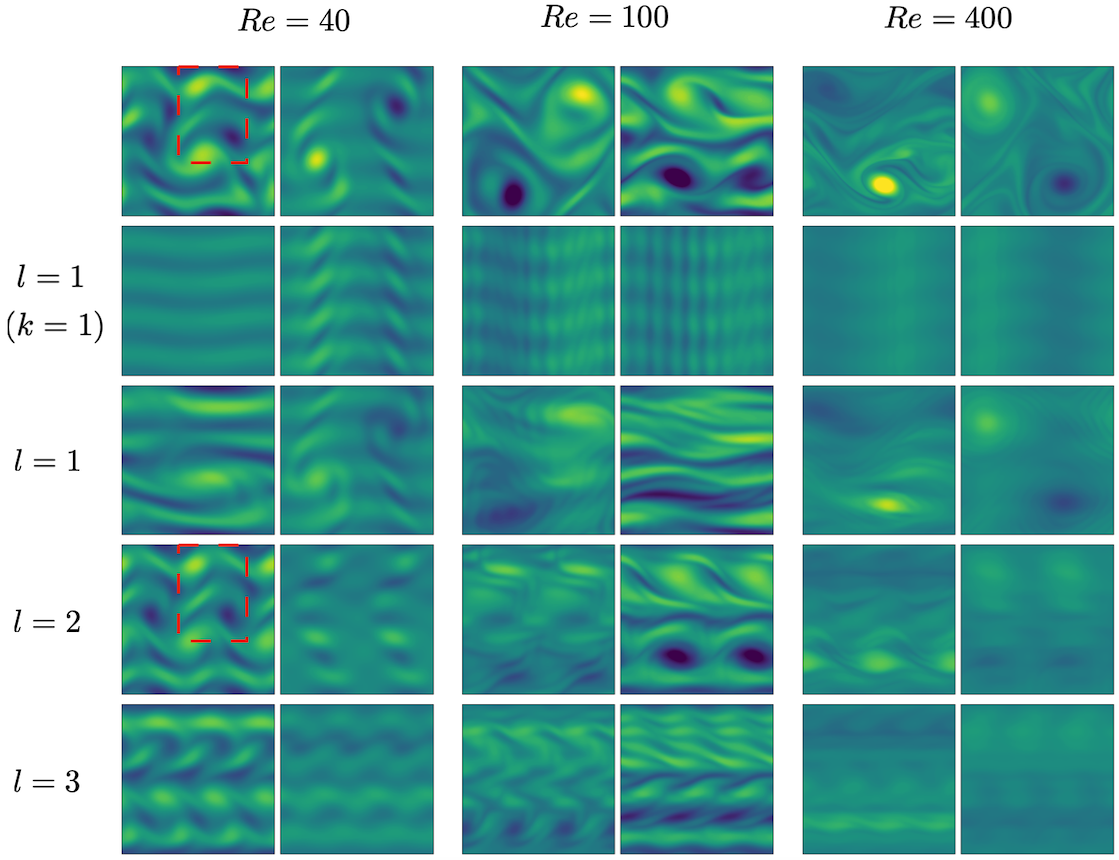}
    \caption{Top: two sample vorticity snapshots each at $Re=40$ (left), $100$ (centre) and $400$ (right). 
    The remaining rows show the decode of the most energetic modes within the $l=0$ and $l=1$ subspace (equation \ref{eqn:most_energetic}) followed by the full decodes of the $l=1$, $l=2$ and $l=3$ projections (equation \ref{eqn:full_latent_decode}).
    At $Re=40$ the left snapshot is an example of a higher dissipation event; the right snapshot at $Re=100$ is also high dissipation.
    Red box highlights a feature discussed in the text. 
    %\rk{[$l=1(0)$ label is confusing, what about $l \in \{0,1\}$ or $l =0\, \& \,1$? or even $\{-l,0,l\}$?]} % JP have adjusted to use the l, k terminology in the text 
    }
    \label{fig:latent_projections}
\end{figure}
% Briefly review idea of decoding individual latent wavenumbers. 
% Argue for (1) dynamical relevance based on established similarity to known equilibria; (2) utility in understanding bursts. 
The physical structures which are associated with a particular latent wavenumber can be identified by decoding projections of snapshots onto individual, or combinations of, latent vectors within that particular eigenspace.
To do this, we follow the methodology outlined in PBK21 and perform an SVD within each eigenspace to obtain a set of mutually orthornormal `patterns', which are ordered by their contribution to the total variance within that particular value of $l$ over the test dataset, and which can be decoded in isolation. 

To perform the SVD within a particular latent subspace $\pm l$, we first project down the latent space to the subspace as follows
%we first construct a pair of biorthogonal bases from the numerically computed left- and right-eigenvectors associated with wavenumber $l$. 
%We then define $\mathscr P^l$ as the projector onto \rk{the whole of} subspace $l$:
\begin{equation}
    \mathscr P^l (\mathscr E (\omega)) = \sum_{k=1}^{d(l)} \mathscr P^l_k (\mathscr E (\omega)) + \text{c.c.},
\end{equation}
%where $(\boldsymbol \xi_i^{\dagger})^H\boldsymbol \xi_j = \delta_{ij}$ are the biorthogonal embedding vectors. \rk{[maybe $\boldsymbol \xi_j^{(l)}?$ see later with u...]}.
Adding the complex conjugate is required when $l \neq 0$. %  \rk{[is this because you are really considering $l$ and $-l$?]}. % yes, adjusted text above 
We then construct the projected embedded data matrix and perform an SVD within that subspace:
\begin{equation}
    \mathbf E^l := \begin{bmatrix}
        \mathscr P^l(\mathscr E(\omega_1)) & \cdots & \mathscr P^l(\mathscr E(\omega_1)) 
        \end{bmatrix} = \mathbf U_l \boldsymbol \Sigma_l \mathbf W_l^H.
        \label{svd}
\end{equation}
The results of decoding various combinations of latent projections within different subspaces are reported in figure \ref{fig:latent_projections}. 
Here we take a pair of snapshots at each $Re \in \{40, 100, 400\}$, compute the embedding $\mathscr E(\omega)$, project onto the individual eigenspaces with $\mathscr P^l$ before decoding some combination of projections, in some cases using only a subset of the left-singular vectors $\{\mathbf u_l^k\}_{k=1}^{d(l)}$ to decode the most energetic components. 
The net result of all this  is  to replace the original basis $\{ \boldsymbol \xi^{(l)}_k\}_{k=1}^{d(l)}$ on the $l$-latent subspace with a better-designed basis $\{\mathbf u_l^k\}_{k=1}^{d(l)}$ based on energy content.

For physically realistic outputs we always include the streamwise-invariant projection onto $l=0$ see discussion in PBK21), and in figure \ref{fig:latent_projections} we consider the combinations
\begin{equation}
    \tilde{\omega} = \mathscr D \left([(\mathbf u_{l=0}^1)^H \mathscr P^{l=0}(\mathscr E(\omega))\mathbf u_{l=0}^1] + [(\mathbf u_{l=1}^1)^H \mathscr P^{l=1}(\mathscr E(\omega))\mathbf u_{l=1}^1 + \text{c.c.}]\right),
    \label{eqn:most_energetic}
\end{equation}
% \rk{[This doesn't look quite right - argument for decode is a scalar?)]}
along with decodes of full eigenspaces
\begin{equation}
    \tilde{\omega} = \mathscr D \left(\mathscr P^{0}(\mathscr E(\omega)) + [\mathscr P^{l}(\mathscr E(\omega)) + \text{c.c.}]\right),
    \label{eqn:full_latent_decode}
\end{equation}
for $l=1, 2, 3$.
In the former case (\ref{eqn:most_energetic}) we project onto only the most energetic directions within $l=0$ and $l=1$ before decoding. 
The most energetic $l=0$ mode decodes by itself to something resembling the basic laminar solution (not shown -- eight bands of vorticity, invariant under all symmetries).
The addition of the most energetic $l=1$ mode results in something that at $Re=40$ closely resembles the first non-trivial equilibrium to bifurcate from the laminar solution at $Re \approx 10$ \citep[see e.g.][PBK21]{Chandler2013}. 
This is visible in the second row of snapshots in figure \ref{fig:latent_projections}.
Notably a similar structure is found at both $Re=100$ and $Re=400$, though it is much less clear (particularly note the high-wavenumber contamination at $Re=100$). 
This reflects the diminishing role of this unstable solution in the dynamics, and also coincides with a relative increase in the energy of other singular vectors across the $l=1$ subspace as $Re$ increases (not shown). 

Decodes of full subspaces obtained via (\ref{eqn:full_latent_decode}) are also reported below the snapshots in figure \ref{fig:latent_projections}. 
% Points to make: doubly/triply periodic from localised features in original snapshot -- will connect later to ismple invariant solutions
% Murky effect at Re=400 not unlike a low-pass filter
The project-and-decode operation (equation \ref{eqn:full_latent_decode}) produces vorticity fields with a fundamental horizontal lengthscale $2\pi / l$, which is set by the latent wavenumber. 
When the flow is dominated by a pair of opposite-signed vortices, the full $l=1$ decode looks visually similar to the input snapshot at the lower $Re=40$, while
for the higher values of $Re=100$ and $Re=400$ the result appear similar to a low-pass filter applied in Fourier space. 
Notably the higher dissipation events (leftmost snapshot at $Re=40$; right snapshot at $Re=100$) do not exhibit this behaviour.
In these cases, the $l=1$ pattern extracts some of the features in the input image, though there is a much stronger response in $l=1$ at $Re=100$ compared to $Re=40$ (discussed further below in \S\ref{sec:high_diss}). 

The $l=2$ and $l=3$ decodes of the higher dissipation events at $Re=40$ and $Re=100$ are particularly interesting, since they highlight local vortical features that make up part of larger, doubly-or-triply periodic pattern within the autoencoder -- for example note the structure identified with the red box at $Re=40$ which then forms the basis for the $l=2$ recurrent pattern. 
The dynamical relevance of these patterns will be explored later when we use them to find new, high dissipation UPOs, which are visually very similar to the $l=2$ (and, at $Re=40$, $l=3$ too) decodes shown here.

% There are several things to note in this figure (presumed, some known TBC as I generate figures)
% \begin{enumerate}
%     \item Laminar solution at all $Re$
%     \item First non-trivial equilibrium at all $Re \neq 400$
%     \item $l=2$ decodes showing interesting structure; might be two figures. 
% \end{enumerate}

\subsection{High dissipation events under increasing $Re$}
\label{sec:high_diss}
%
% Fig 7
%
\begin{figure}
    \centering
    \includegraphics[width=0.32\textwidth]{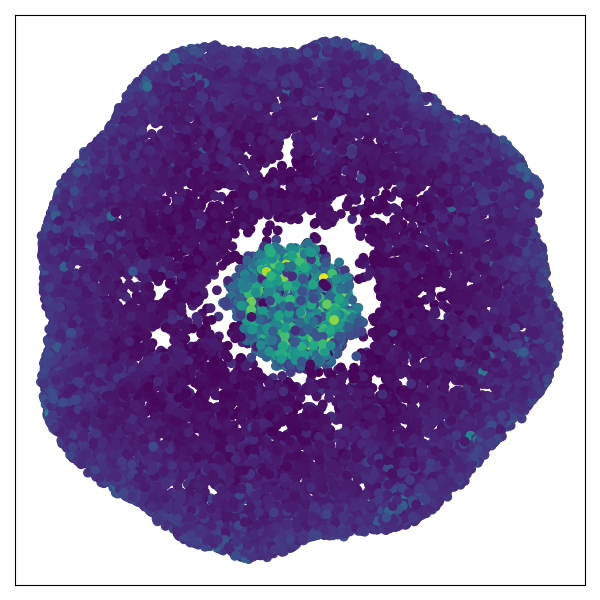}
    \includegraphics[width=0.32\textwidth]{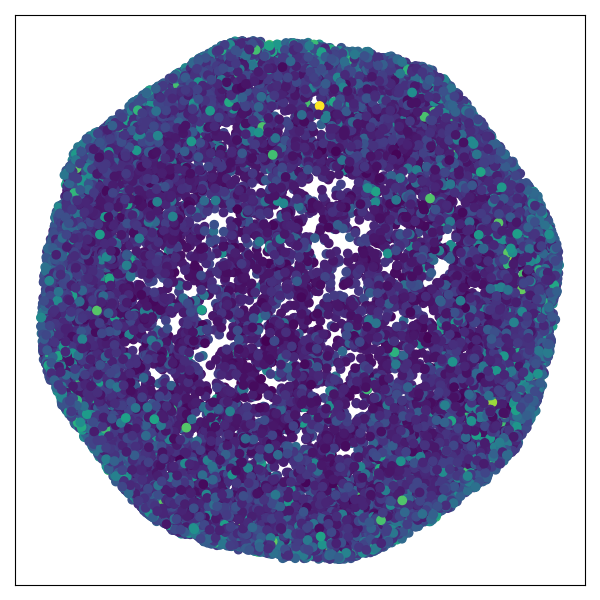}
    \includegraphics[width=0.32\textwidth]{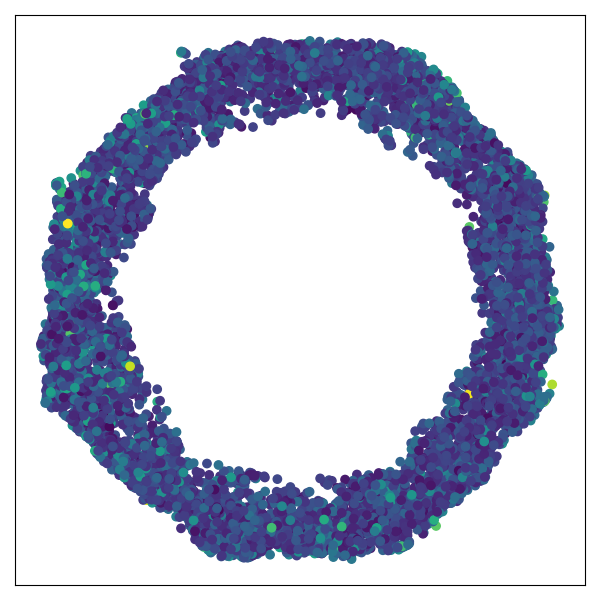}
    \caption{
    Two-dimensional visualisations of the continuous-shift invariant observable (\ref{eqn:latent_observable}) obtained using the UMAP algorithm \citep{2018arXivUMAP}.
    From left to right are the results for $Re=40$ ($m=128$ network), $Re=100$ ($m=512$ network) and $Re=400$ ($m=1024$ network). 
    Note the octagonal shape visible in all three examples represents the networks' internal representation of the discrete shift-reflect symmetry in the system.
    Data points are coloured by their dissipation values running from the lowest (dark blue) to highest (yellow). 
    }
    \label{fig:burst_umap}
\end{figure}

The structure of the state space of vorticity fields can be explored further by dimensionality reduction on latent Fourier projections. 
We first define a streamwise-shift invariant observable using the latent wavenumbers $0 \leq l \leq 3$,
\begin{equation}
    \boldsymbol \psi(\omega) := \begin{pmatrix}
        (\mathbf u_{l=0}^1)^H \mathscr P^{0}(\mathscr E(\omega)) \\
        (\mathbf u_{0}^2)^H \mathscr P^{0}(\mathscr E(\omega)) \\
        \vdots \\
        |(\mathbf u_{1}^1)^H \mathscr P^{1}(\mathscr E(\omega))| \\
        \vdots \\
        |(\mathbf u_{3}^{d(3)})^H \mathscr P^{3}(\mathscr E(\omega))|
    \end{pmatrix},
    \label{eqn:latent_observable}
\end{equation}
where the absolute value for wavenumbers $l>0$ removes any $x$-location dependence of the features and $\mathbf u_l^j$ is the $j^{\text{th}}$ left-singular vector from the matrix $\mathbf U_l$ (defined in (\ref{svd})). 
We then input the streamwise-shift-independent latent observable (\ref{eqn:latent_observable}) into the `UMAP' algorithm \citep{2018arXivUMAP}, which seeks a two-dimensional representation of the data by (i) assuming there is some manifold on which the data is uniformly distributed and (ii) attempting to preserve geodesic distances on the manifold in the Euclidean distances between points in the mapped representation. 
Note we obtain similar results (not shown) with the t-SNE algorithm \citep{tsne}.

Low-dimensional visualisations produced by the combination of equation (\ref{eqn:latent_observable}) and the UMAP algorithm are reported in figure \ref{fig:burst_umap} for all three networks examined in detail.
In each case there is a clear octagonal shape in the data cloud, which is due to the shift-reflect symmetry present in the full equations (\ref{eqn:vorticity_kf}): i.e. each of the eight sectors of the octagon is, to a good approximation, a shift-reflected copy of the others. 
Increasing the Reynolds number fundamentally alters the latent representations, which allows us to infer something about the nature of the inertial manifold of the governing equations.
At the lowest value of $Re=40$, there is a small detached high-dissipation octagon clearly visible in figure \ref{fig:burst_umap} (a similar structure was observed in PBK21 at $Re=40$, but without the clean shift-reflect embedding due to the poor performance of that autoencoder model). 
This detachment suggests that high dissipation events at $Re=40$ are distinct from the low-dissipation dynamics, which are built around the first non-trivial equilibrium discussed around figure \ref{fig:latent_projections} above. 
% This is further emphasised in the time series reported in figure \ref{fig:latent_burst}, which shows bursting (high dissipation) events to be dominated by recurrent patterns at $l=2$ and $l=3$, with only a small contribution from larger-scale $l=1$ structure. 

In contrast, the low-dimensional visualisation at $Re=100$ in figure \ref{fig:burst_umap} shows no detached bursting structure, but rather the high dissipation events are included in the single embedding `octagon'. 
The embeddings still contain the signature of the non-trivial first equilibrium (although now much more weakly -- see figure \ref{fig:latent_projections}) and there is then a continuous latent connection to any vorticity field regardless of the strength of the dissipation.
More concretely, consider the embedding of a vorticity field and an equivalent, shift-reflected version, $\mathscr E(\omega)$ and $\mathscr E(\mathscr S^m \omega)$ ($1 \leq m \leq 7$): 
it is possible to reach the embedding $\mathscr E(\mathscr S^m \omega)$ from $\mathscr E(\omega)$ without having to go through intermediate shift reflects 
-- i.e. any shift-reflected copy can be smoothly reached by passing through the middle of the octagon. 
% This is supported by the results in figure \ref{fig:latent_burst}, which continue to indicate the strong role of large, $l=1$-scale structure when high dissipation dynamics are recorded. 
% High-dissipation events coincide with the emergence of a local small scale structure, in contrast to the emergence of small scale structures throughout the domain at $Re=40$.

%
% FIG 8
%
\begin{figure}
    \centering
    \includegraphics[width=0.32\textwidth]{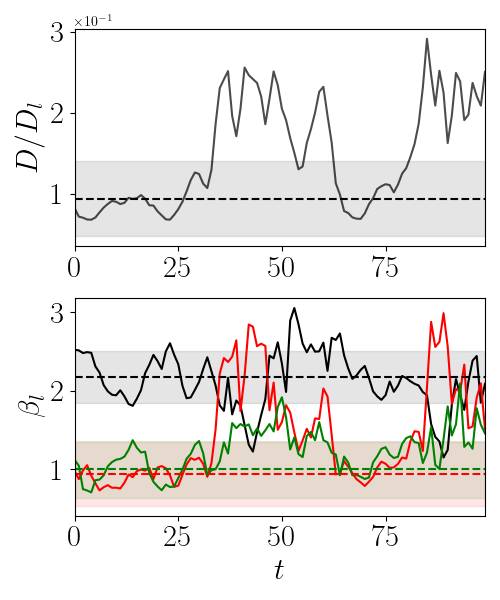}
    \includegraphics[width=0.32\textwidth]{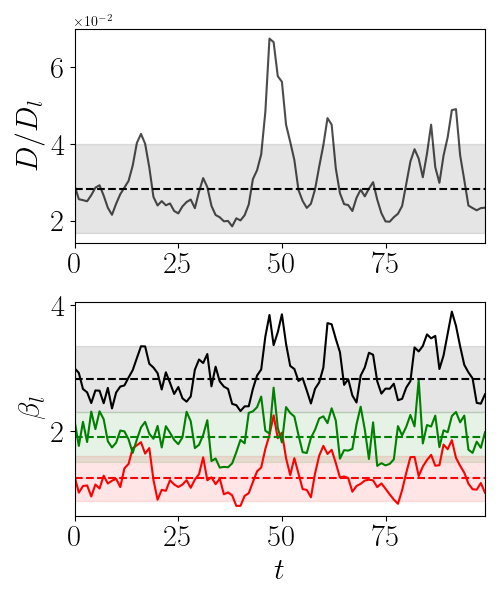}
    \includegraphics[width=0.32\textwidth]{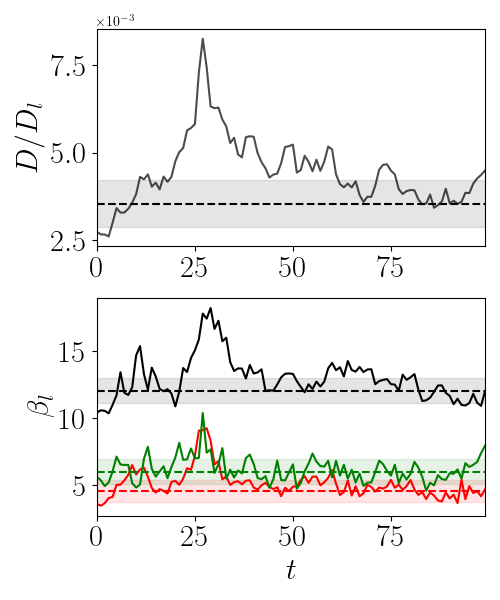}
    \caption{
    Visualisation of individual trajectories at $Re=40$ (left), $Re=100$ (middle) and $Re=400$ (right). 
    Top panels show dissipation rate normalised by the laminar value, while the lower panels show the quantity $\beta_l$ (equation \ref{eqn:burst_flag}) for $l=1$ (black), $l=2$ (red) and $l=3$ (green).
    In all cases the dashed horizontal lines are the mean value of the quantity (sample mean over the test set), while the shading identifies $\pm 1.5$ standard deviations. 
    % Visualisation of bursts as projections onto various latent wavenumbers. Should emphasize role of $l=1$ when $Re>40$, perhaps some clarity for $Re=400$.
    }
    \label{fig:latent_burst}
\end{figure}
The lower-$Re$ embeddings should be contrasted with the results at $Re=400$, which indicate an entirely different latent representation. 
The octagonal shape remains in figure \ref{fig:burst_umap}, but now with a large, central hole.  %-- this gap coincides with the absence of the primary non-trivial equilibrium which could be observed at the lower $Re$ values. 
It is now only possible to reach a shift-reflected copy of the embedding of a vorticity field by traversing around the octagonal ring -- moving incrementally through shift-reflects.
% Instead, the network has built a representation based on the continued presence of a large domain-filling vortex pair. 
This reflects the fact that the representations are now built around the continued presence of a large domain-filling vortex pair, and coincides with the diminished importance of the first non-trivial equilibrium which is so essential to the dynamics at lower $Re$. 

% The time series of the high-$Re$ network reported in figure \ref{fig:latent_burst} supports these assertions, 
% COMMENTS ON TIME SERIES. 
To further explore the changing nature of `high dissipation' events as $Re$ increases, we compute the following observable for individual trajectories in our test set:
\begin{equation}
    \beta_l(\omega) := \sum_{k=1}^{d(l)}|(\mathbf u_{l}^{k})^H \mathscr P^{l}(\mathscr E(\omega))|,
    \label{eqn:burst_flag}
\end{equation}
which is a measure of the total contribution of the latent subspace $l$.
We compare the evolution of (\ref{eqn:burst_flag}) to the instantaneous dissipation rate in figure \ref{fig:latent_burst} for wavenumbers $l\in \{1,2,3\}$, where we also indicate the mean value of each quantity over the entire test set for a meaningful comparison.

For the trajectory at $Re=40$ in figure \ref{fig:latent_burst} two high-dissipation events are observed (both are substantially longer than typical excursions from the low dissipation dynamics). 
In both cases, the burst is associated with extreme values (2-3 standard deviations above the mean) in latent wavenumbers $l=2$ and $l=3$, as measured by equation (\ref{eqn:burst_flag}). 
Note also that this amplification coincides with a relatively low value of the amplitude in $l=1$. 
The earlier UMAP projections are consistent with this: to a reasonable approximation the detached burst octagon in figure \ref{fig:burst_umap} reflects that low-dissipation events are primarily $l=1$, while high dissipation is largely associated with $l=2$ and $l=3$ patterns. 

Similarly, the high dissipation event observed at $t \sim 50$ for the $Re=100$ trajectory in figure \ref{fig:latent_burst} is also associated with a local spike in the $l=2$ contribution, and a smaller jump in $l=3$.
However, the burst also features larger-than-usual amplitude in $\beta_1$.
Bursting events feature small scale vortices (corresponding to $l=2$ and $l=3$ patterns), but \emph{locally}, and large-scale $l=1$ structure continues to play a role -- hence there is no detached burst octagon in figure \ref{fig:burst_umap}.
This effect is amplified further at $Re=400$, where the high dissipation values recorded at $t\sim 25$ in figure \ref{fig:latent_burst} are now associated with extreme values in the largest scale patterns at $l=1$.
% There is a limited role for localised, small-scale structures in this flow. 
% CONNECT TO EULER SOLUTIONS COMING DOWN GRIGORIEV ETC. SPECULATE.

The merging and subsequent disappearance of distinct high dissipation events described in figure \ref{fig:burst_umap} can be connected to small-scale UPOs and their movement away from the attractor as $Re$ increases. 
To explore these effects, we now use our embeddings to generate UPO guesses in two ways, both by modifying the classical approach \citep{Kawahara2001, Viswanath2007, Cvitanovic2010,Chandler2013} and by using individual latent wavenumbers to look for high dissipation solutions.

\section{Unstable periodic orbits}
\label{sec:upos}
\subsection{Recurrent flow analysis with latent variables at $Re=40$}
The classical method for searching for UPOs in a turbulent flow is to measure similarity between vorticity fields on the same orbit, separated by $T$ in time. 
If the vorticity field is `similar' now to a point $T$ in the past, this is taken to indicate that the flow has shadowed a UPO for a full period $\sim T$. 
To find the UPO, the starting (past) vorticity field is input into a Newton-Raphson algorithm, along with the guess for the period, $T$.
This is `recurrent flow analysis' \citep{Kawahara2001,Viswanath2007,Cvitanovic2010,Chandler2013}, and the `similarity' between snapshots is measured via a Euclidean norm:
\begin{equation}
    R(\omega, T) := \min_{s}\frac{\| \mathscr T^{s} f^T(\omega) - \omega \|}{\|\omega\|},
    \label{eqn:rfa_norm}
\end{equation}
where $f^T$ is the time-forward map of equation (\ref{eqn:vorticity_kf}) and we perform a search over the continuous symmetry (note we could also search over the discrete symmetries, but we do not consider this here). 
Local minima in $R(\omega,T)$ which fall below some threshold are selected as viable guesses for UPOs which take the form of a triple $(\omega, T, s)$ (which includes a guess for the shift $s$ that minimises the right hand side of \ref{eqn:rfa_norm}).
In Kolmogorov flow a relatively large value of $R$ is required to flag guesses \citep[e.g. usually a relative error of $R=0.3$ or higher is used, see][]{Chandler2013,Lucas2014,Lucas2015}.
This is similar to the the issues discussed around the error metric (\ref{eqn:local_err}) used for the neural networks in \S \ref{sec:densenet}. 

There are two shortcomings with the approach outlined above:
(i) it requires a near recurrence to occur, which is unlikely at higher $Re$ due to the increased instability of the UPOs, and
(ii) the measure in equation (\ref{eqn:rfa_norm}) is unlikely to be the best measure of similarity for points which are not particularly close on the inertial manifold, but which may still be observed in a shadowing event. 
At $Re=40$ recurrent flow analysis can still be somewhat effective, hence we focus on point (ii) here.

The central idea is that the features recorded in our autoencoders are likely to be a much more effective observable for flagging similarity between snapshots than the computational vector of the vorticity field itself. 
We therefore keep the main mechanics of a recurrent flow analysis in place, but instead measure similarity in the shift-independent observable $\boldsymbol{\psi}(\omega)$ defined in equation (\ref{eqn:latent_observable}).
Our modified near-recurrence function reads
\begin{equation}
    R_{\mathscr E}(\omega, T) := \frac{\| \boldsymbol \psi(f^T(\omega)) - \boldsymbol \psi(\omega) \|}{\|\boldsymbol\psi(\omega)\|}.
    \label{eqn:rfa_latent}
\end{equation}
Now no search over the continuous symmetry is required. 
The guess for the shift is determined from the phase difference in the projection onto the dominant $l=1$ mode between the `start' and `end' snapshots via
\begin{equation}
    s = -i \, \log\left(\frac{(\mathbf u_{l=1}^{1})^H \mathscr P^{l=1}(\mathscr E(f^T(\omega)))}{(\mathbf u_{l=1}^{1})^H \mathscr P^{l=1}(\mathscr E(\omega))}\right) .
\end{equation}

%
% Fig 9
%
\begin{figure}
    \centering
     \includegraphics[width=\textwidth]{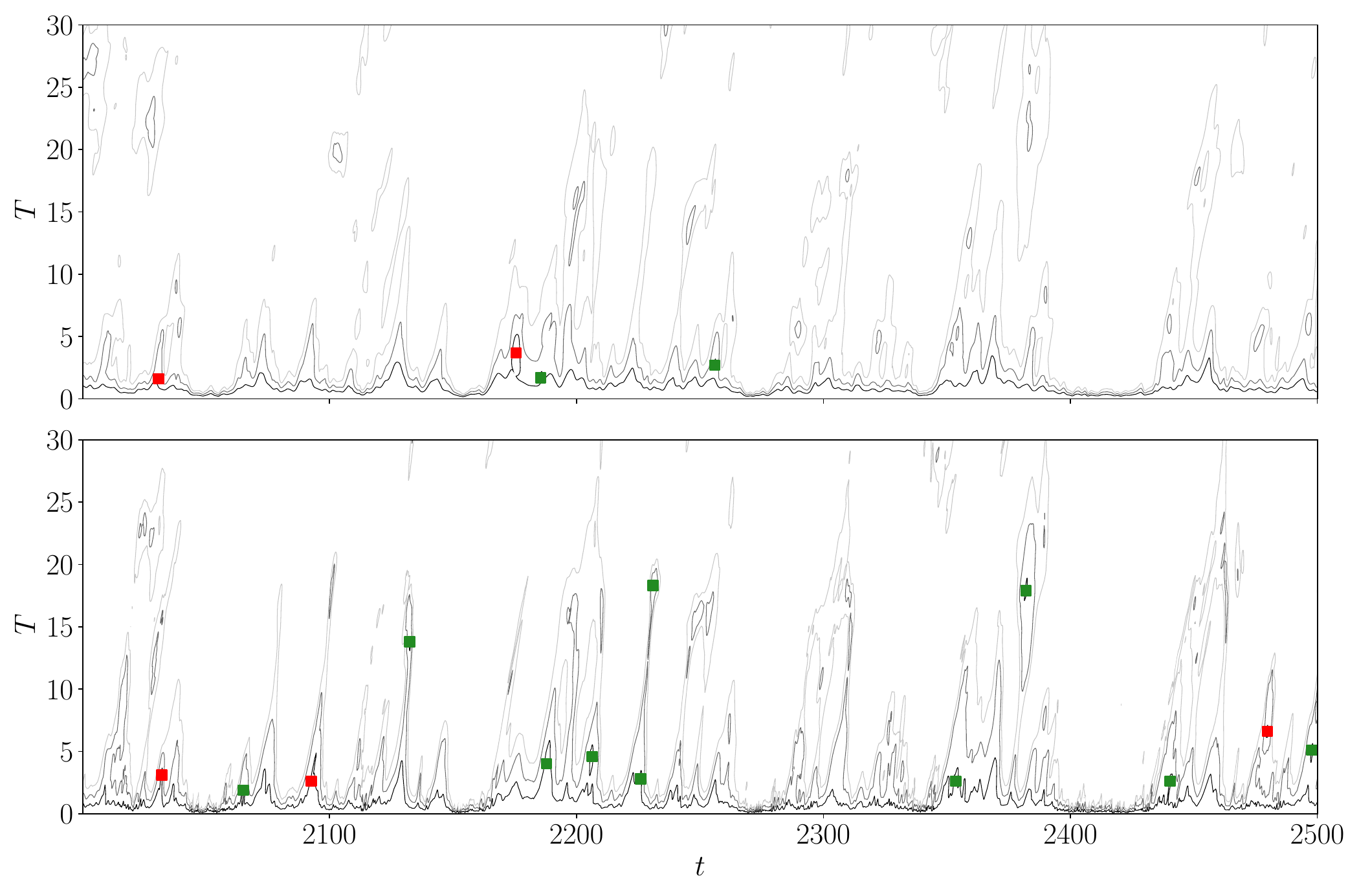}
    \caption{Recurrent flow analysis comparison between the full vorticity field (\ref{eqn:rfa_norm}) and the latent observable (\ref{eqn:rfa_latent}), for the same example time series. 
    (Top) Contours of recurrence measure $R=0.3$ (black) $R=0.45$ (dark grey) and $R=0.6$ (lightest grey). Points where $R\leq 0.3$ were supplied as initial guesses to a Newton solver. 
    (Bottom) Contours of latent recurrence measure $R_{\mathscr E} = 0.01$ (black), $R_{\mathscr E} = 0.02$ (dark grey) and $R_{\mathscr E} = 0.03$ (lightest grey). Points where $R_{\mathscr E} \leq 0.01$ were supplied as initial guesses to a Newton solver (selection of this threshold is discussed in the main text).  
    %$R$ contours at 0.3, 0.45 and 0.6. $R_{\mathscr E}$ contours at 0.01, 0.02, 0.03. For normal RFA there were 282 guesses, 73 of which converged to give 20 unique solutions. For latent RFA there were 405 guesses, 168 of which converged to give 29 unique solutions.
    In both cases red dots indicate a failed convergence, green dots a success. }
    \label{fig:rfa}
\end{figure}

%
% Fig 10
%
\begin{figure}
    \centering
    \includegraphics[width=0.85\textwidth]{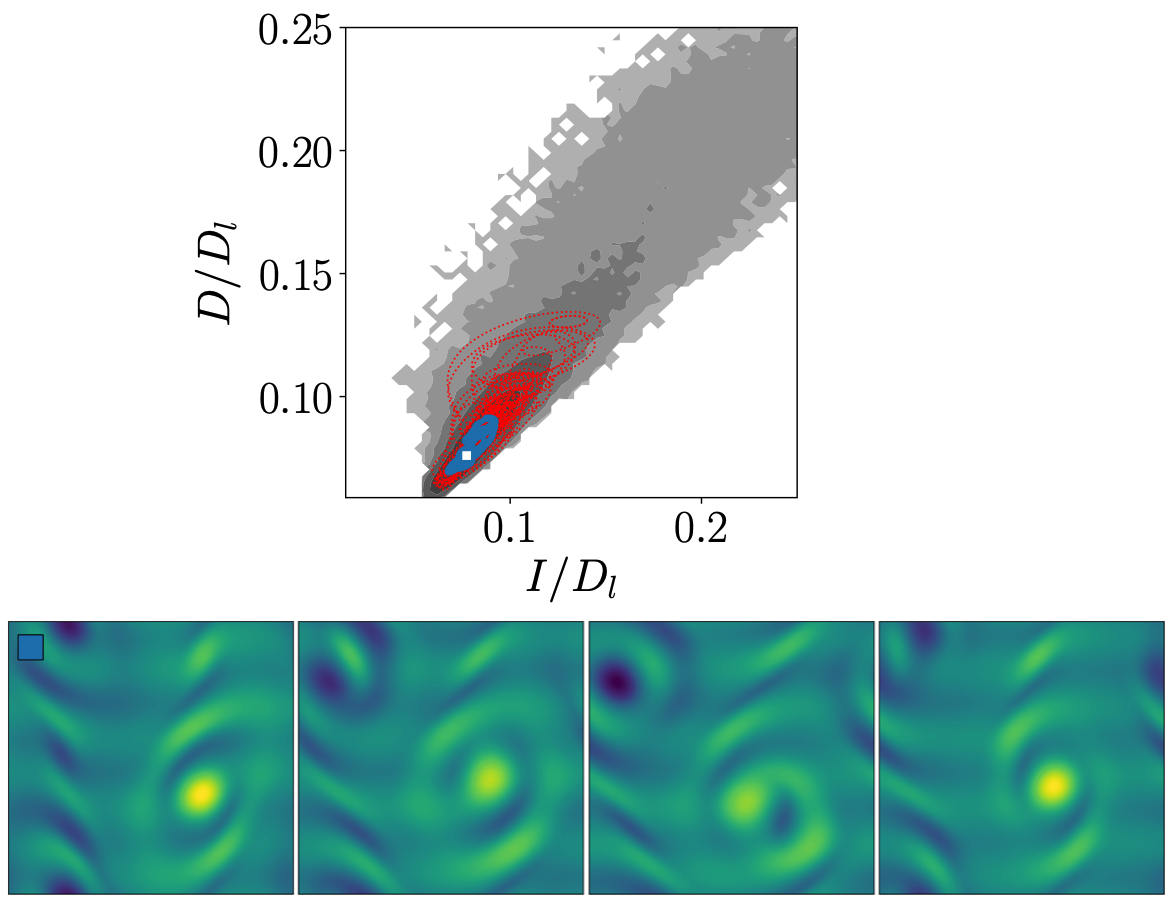}
    \caption{(Top) All periodic orbits found via latent Fourier-based recurrent flow analysis, visualised in terms of their production and dissipation rates (red lines). 
    The laminar dissipation value $D_l = Re/ (2 n^2)$.
    Background grey contours are the turbulent p.d.f. (contour levels spaced logarithmically with a minium value $10^{-6}$), while the blue line identifies the longest UPO found with $T=49.091$, for which we show four snapshots of spanwise vorticity equally spaced every $T/5$ in the lower four panels of the figure. The starting point for the visualisation of the vorticity is identified with a white square in the production/dissipation plot, while contour levels for the vorticity run from $-10 \leq \omega \leq 10$.}
    \label{fig:rfa_pos}
\end{figure}

The performance of the new latent observable is compared to the standard approach in figure \ref{fig:rfa}, where we report contours of the regular recurrence measure $R$ and $R_{\mathscr{E}}$ over $500$ advective time units. 
Following the protocol defined in \cite{Chandler2013}, initial conditions $\omega(t)$ for which $R(\omega(t), T) < 0.3$ are supplied as guesses to a Newton-Krylov-Hookstep solver in an attempt to converge an exact UPO. 
To define when guesses are initialised based on the latent observable, we seek a threshold value of $R_{\mathscr{E}}$ which recovers a substantial portion of the original `standard' guesses based on $R_{\omega}$. A value of $R_{\mathscr{E}}=0.01$ is found to be robust -- and is actually perhaps a rather conservative choice as discussed further below. 

The contour plots reported in figure \ref{fig:rfa} indicate that the flagging of near recurrences is often qualitatively similar between the two approaches. 
The fact that the latent observable flags many of the same guesses is to be expected, but it also identifies near recurrences which are not observed at all in the original vorticity-based approach, even when the recurrence threshold is relaxed. 
In figure \ref{fig:rfa} we have also included contours of $R(\omega, T)$ significantly above the threshold (as high as a relative error of 60\%). 
We can see that the latent observable does indeed identify some UPOs that were missed because they were just above the standard vorticity threshold, for example note the convergences around $t\sim 2200$, but also identifies near recurrences which would not have been identified at all with the previous approach (note the success with $T \sim 20$ at $t \sim 2230$). 

We applied both standard and `latent' recurrent flow analysis over a single trajectory of length $0 \leq t \leq 8000$, with a maximum period in the search of $T_{\text{max}}=50$. 
The standard approach resulted in 232 guesses, of which 73 UPOs were converged, with 20 unique solutions -- a success rate of $\sim 26 \%$ in terms of raw convergences.
In contrast, the latent Fourier approach produced a much larger number of 405 guesses, of which 168 converged to UPOs, with 29 unique solutions.
In the latter case, there is a much higher convergence rate of $\sim 42 \%$, which indicates that the threshold on $R_{\mathscr E}$ could be relaxed considerably. 
Even with the relatively conservative choice of $R_{\mathscr E} = 0.01$, we have still more than doubled the raw number of convergences in a long time series. 

The periodic orbits found using the latent recurrent flow analysis are shown in figure \ref{fig:rfa_pos} in the form of two-dimensional projections onto a production-dissipation-rate diagram which also includes the turbulent p.d.f. 
Similar to the majority of structures found previously via recurrent flow analysis \citep[e.g. in particular see][]{Chandler2013,Lucas2014}, all the UPOs found are relatively low dissipation. 
So, while latent recurrent flow analysis does provide access to large numbers of new solutions that a standard recurrent flow analysis has not been able to return, it is still constrained by the fact that the more unstable structures are not flagged in this approach at all. 
We now discuss a new method to use latent Fourier analysis to isolate smaller-scale solutions which play a substantial role in the high-dissipation dynamics, and which is effective at both $Re=40$ and $Re=100$. 

\subsection{Bursting periodic orbits}
In PBK21 we used projections onto the $l=2$ latent Fourier modes to find large numbers of equilibria and travelling waves which bore some resemblance to snapshots of high dissipation events at $Re=40$. 
The quality of the networks constructed here allows us to go much further and find large numbers of high dissipation UPOs.
Our method is also effective at $Re=100$, though the structures we isolate there play a slightly different role in the dynamics. 

We have seen that the latent observable (\ref{eqn:latent_observable}) can substantially improve the performance of recurrent flow analysis, though as described above the new UPOs identified in this way are largely `low dissipation' (all have roughly $D/D_{lam} \lesssim 0.15$). 
An alternative approach is motivated by the time series in figure \ref{fig:latent_burst} which show a correlation of high dissipation events with projections onto $l=2$ and $l=3$ latent recurrent patterns. 
These projections indicate the role of smaller-scale structure in the bursts and we therefore search for UPOs with a smaller fundamental horizontal lengthscale (i.e. $L_x /2$ or $L_x / 3$).

We generate guesses for these small-scale, high dissipation UPOs by decoding the projection onto either $l=2$ or $l=3$ only (inclusion of the $l=0$ subspace is always required for a physically realistic output, see discussion in PBK21):
\begin{equation}
    \omega_g = \mathscr D\left(\mathscr P^{0}(\mathscr E(\omega)) + \left[\mathscr P^{l}(\mathscr E(\omega)) + \text{c.c.}\right] \right).
    \label{eqn:po_burst_guess}
\end{equation}
We create UPO guesses by evaluating (\ref{eqn:po_burst_guess}) on snapshots $\omega$ for which the quantity (\ref{eqn:burst_flag})
% \begin{equation}
%     \beta_l(\omega) := \sum_{k=1}^{d(l)}|(\mathbf u_{l}^{k})^H \mathscr P^{l}(\mathscr E(\omega))|,
%     \label{eqn:burst_flag}
% \end{equation}
was greater than two standard deviations above the mean. 
We uniformly set the initial guess for the period to $T=5$ at $Re=40$ and $T=2.5$ at $Re=100$, values motivated by the typical duration of a `bursting' event. 
The initial guess for the shift was set to zero throughout, $s=0$.  

%
% FIG 11
%
\begin{figure}
    \centering
    \includegraphics[width=0.4\textwidth]{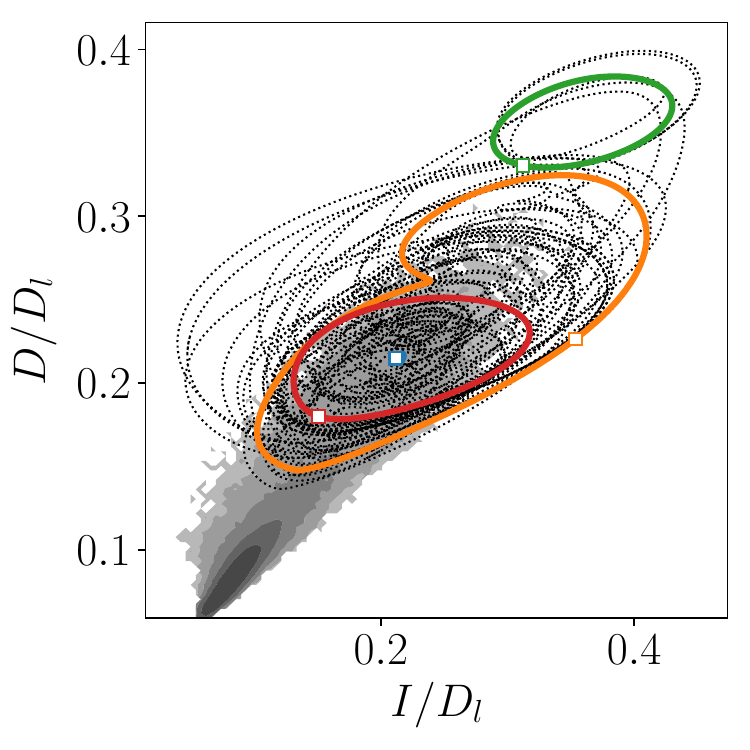}
    \includegraphics[width=0.57\textwidth]{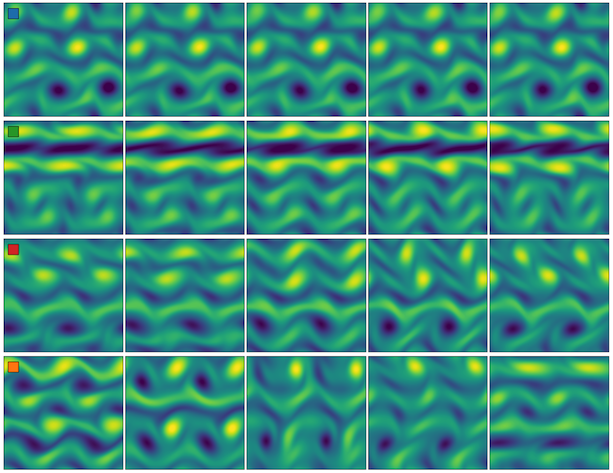}
    \caption{(Left) Production/dissipation PDF of a long turbulent trajectory (contour levels spaced logarithmically with a minium value $10^{-6}$), with all 61 bursting periodic orbits found from $l=2$ projections at $Re=40$ (dotted lines). %\rk{[just a thought: have we defined $D_l$ anywhere?...here $l$ means `laminar' rather than latent wavenumber $l$ of course...]} 
    The laminar dissipation value $D_l = Re/ (2 n^2)$. 
    Four periodic orbits are highlighted in color. Blue: $T=3.962$ (note the very small loop can't be seen clearly in the visualisation on the left); green: $T=2.606$; orange: $T=2.895$ and red: $T=7.452$. (Right) Spanwise vorticity at points equispaced in time for the four highlighted UPOs, with $\Delta t = T/5$ and starting at the point indicated by the relevant marker in the left panel. Vorticity contours run from $-10 \leq \omega \leq 10$ (though the maximum vorticity in the snapshots is higher than this). }
    \label{fig:burst_Re40_POs}
\end{figure}
The results of the bursting search at $Re=40$ with $l=2$ are summarised in figure \ref{fig:burst_Re40_POs}, where we display all UPOs identified using equation (\ref{eqn:po_burst_guess}) in a production-dissipation plot.
We find 61 unique UPOs, all of which are `high dissipation' and have not been returned by any previous search method \citep[for context, the search using recurrent flow analysis by][over $10^5$ time units resulted in 51 unique low dissipation UPOs]{Chandler2013}.

The 61 $l=2$ UPOs found at $Re=40$ involve various flow structures and can be split into groups with common vortex dynamics -- some of these are highlighted in the panels on the right of figure \ref{fig:burst_Re40_POs}. 
The examples shown in the figure include near-static vortex crystals with small motion of an isolated vortex, a strong band of negative vorticity that rapidly advects the weaker structures around it, corotating pairs of vortices and also the emergence of dipole structures (see final example in figure \ref{fig:burst_Re40_POs}).
Note that all POs found using the $l=2$ projections are exactly, or nearly exactly, doubly-periodic in the streamwise direction, which is expected given the projection onto $l=2$ for the initial guess.
This highlights the role of small-scale vortical structures in the burst and the relevance of smaller-scale UPOs in describing these events. 

%
% FIG 13
%
\begin{figure}
    \centering
    \includegraphics[width=0.4\textwidth]{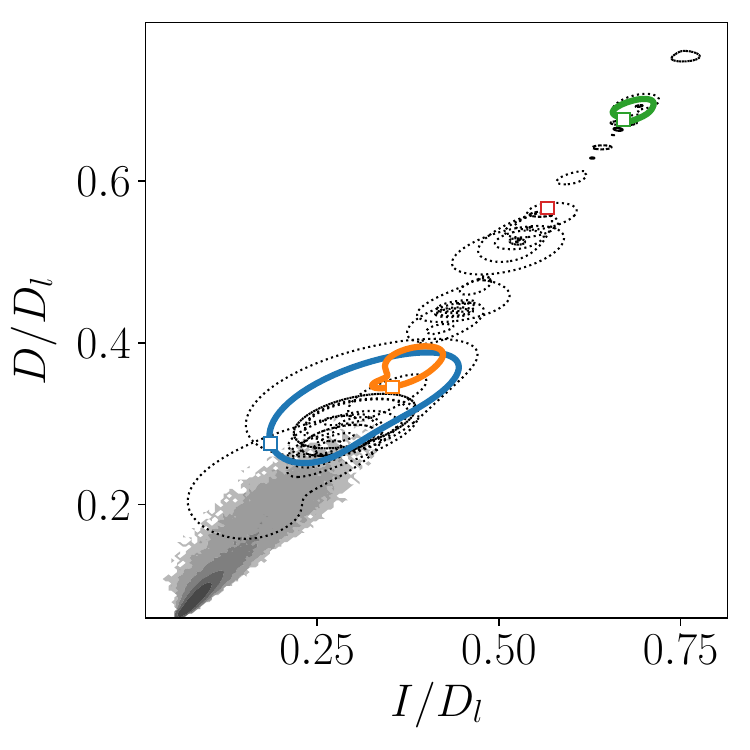}
    \includegraphics[width=0.57\textwidth]{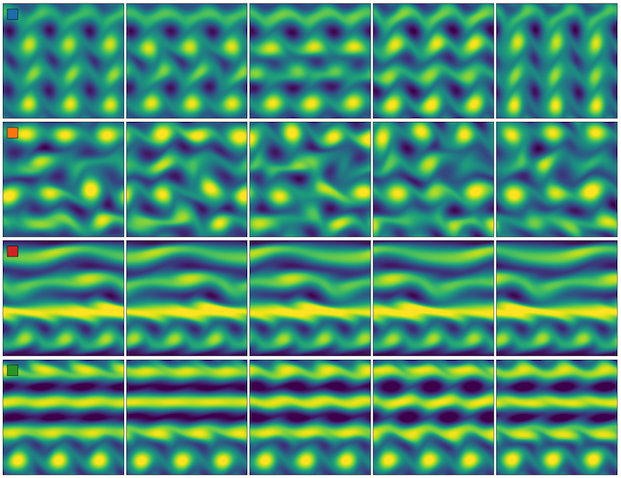}
    \caption{(Left) Production/dissipation PDF of a long turbulent trajectory (contour levels spaced logarithmically with a minium value $10^{-6}$), with all 43 bursting periodic orbits found from $l=3$ projections at $Re=40$ (dotted lines). 
    The laminar dissipation value $D_l = Re/ (2 n^2)$.
    Four periodic orbits are highlighted in color. Blue: $T=5.521$; orange: $T=5.548$; red: $T=2.636$ (note the very small loop can't be seen clearly in the visualisation on the left) and green: $T=3.456$. (Right) Spanwise vorticity at points equispaced in time for the four highlighted UPOs, with $\Delta t = T/5$ and starting at the point indicated by the relevant marker in the left panel. Vorticity contours run from $-10 \leq \omega \leq 10$ (though the maximum vorticity in the snapshots is higher than this). }
    \label{fig:burst_l3_Re40_POs}
\end{figure}

The search using projections onto $l=3$ was almost as prolific, and resulted in $43$ high-dissipation periodic orbits which are summarised in figure \ref{fig:burst_l3_Re40_POs}. 
While these smaller-scale vortical structures play a role in the high dissipation events -- as evidenced by the strong $l=3$ projections in figure \ref{fig:latent_burst} -- they tend to occur alongside other larger scale features, typically associated with patterns dominated by $l=2$.  
The production-dissipation rate projections of the $l=3$ UPOs reported in figure \ref{fig:burst_l3_Re40_POs} are consistent with this -- they almost all sit at much higher dissipation values than those ever observed in a turbulent simulation, simply because they tend to feature very large numbers of high-amplitude vortices. 
The suggestion is that a structure found in an $l=3$ (triply periodic) UPO is observed locally in the true turbulence alongside other larger scale structure. 

% NB is APS abstract (2019!) the best ref for this -- I can't find anything else/more recent from Predrag, could cite Gudorf thesis  - have sent an email to Predrag...Rich

This multiscale, spatiotemporal `tiling' of turbulence by smaller-scale UPOs hinted at here has been explored recently in \citet{Gudorf2019} in the context of simpler dynamical systems, and 
further support for this picture is provided within the $l=3$ solutions explicitly shown in figure \ref{fig:burst_l3_Re40_POs} themselves.
In particular, the third solution (highlighted by the red square) is itself multiscale:
In the top half of the domain there is large $l=1$ dipole structure, while the lower half of the domain is made up of a triply-periodic (three copies of the same pattern) train of vortices. 
The $l=1$ dipole propagates from the right to left, while the train of vortices moves in the other direction, the advection being driven primarily by a pair of opposite-signed, high-amplitude vortex sheets.
Visually, the result looks like the superposition of a pair of counter-propagating travelling waves, with some small time-dependent motion in the vortex sheets. 

%
% FIG 14
%
\begin{figure}
    \centering
    \includegraphics[width=0.4\textwidth]{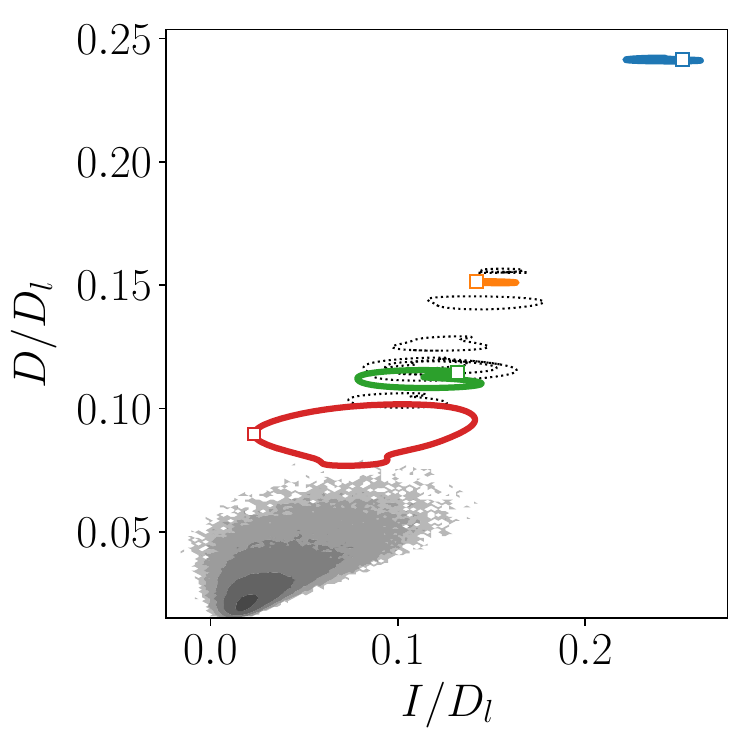}
    \includegraphics[width=0.57\textwidth]{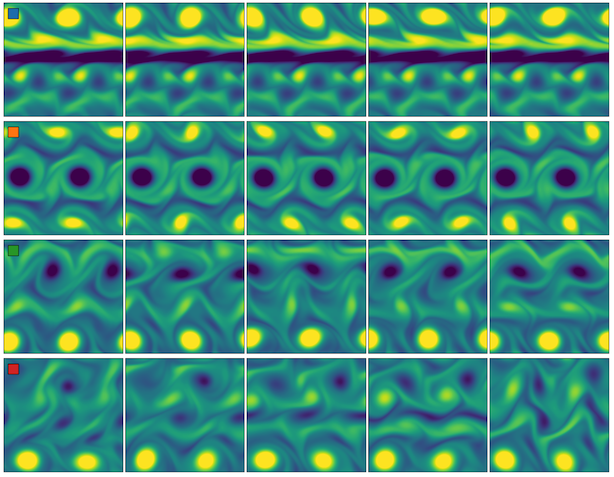}
    \caption{
    (Left) Production/dissipation PDF of a long turbulent trajectory (contour levels spaced logarithmically with a minium value $10^{-6}$), with all 12 bursting periodic orbits found from $l=2$ projections at $Re=100$ (dotted lines). 
    The laminar dissipation value $D_l = Re/ (2 n^2)$.
    Four periodic orbits are highlighted in color. Blue: $T=1.062$; green: $T=1.684$; orange: $T=2.901$ and red: $T=4.499$. (Right) Spanwise vorticity at points equispaced in time for the four highlighted UPOs, with $\Delta t = T/5$ and starting at the point indicated by the relevant marker in the left panel. Vorticity contours run from $-15 \leq \omega \leq 15$ (though the maximum vorticity in the snapshots is higher than this). 
    }
    \label{fig:burst_Re100_POs}
\end{figure}
A similar search was conducted at $Re=100$, resulting in 12 unique high-dissipation UPOs from latent wavenumber $l=2$ (no solutions were found using the $l=3$ projections). 
Two dimensional projections of all 12 $l=2$ solutions are shown in figure \ref{fig:burst_Re100_POs} overlayed on the turbulent p.d.f. 
The most striking departure from the equivalent results at $Re=40$ is that all the converged UPOs have moved away from the (projection of) the attractor: the dissipation values for the UPOs are much higher than those typically observed on a turbulent orbit; the same behaviour as observed with $l=3$ at $Re=40$.  
The reason for is also the same -- we rarely observe dominant $l=2$ structure in realistic turbulent snapshots at $Re=100$ -- for instance notice in figure \ref{fig:latent_burst} the continuing importance of the $l=1$ structures in the high-dissipation events.
Instead, the increased importance of $l=2$ modes coincides with the appearance of smaller-scale, \emph{spatially localised} flow structures. 

% needs work ... 
The snapshots from some of the $Re=100$ UPOs shown in the right panel of figure \ref{fig:burst_Re100_POs} all feature structures that are observed locally in turbulent snapshots, but all have a fundamental streamwise wavelength (or at least nearly) of $\lambda = \pi$. 
% This suggests that the solutions are realised locally in the domain in a realisation of the spatiotemporal tiling described recently by REFs. 
% Comments -- (1) vortices are typically larger -- cf at Re=40 they roughly line up with the 8 forcing bands, here individual 1/4 domain structures are common, (2) dipolar structures found in simple invariant solutions at Re=40 are largely absent. Dynamics feature co-rotating pairs, quasi-static crystals, and the banded structure
In contrast to the $Re=40$ structures reported in figure \ref{fig:burst_Re40_POs}, the vortices in the $Re=100$ UPOs tend to be larger in scale -- some occupying $\sim 1/4$ of the full domain -- while the lower $Re$ solutions still show the signature of the forcing in the vorticity equations. 
The dipoles seen at $Re=40$ are also largely absent in the $Re=100$ UPOs, which tend to be quasi-static crystals, or feature co-rotating pairs of vortices.
The highest-dissipation state features a large band of vorticity which advects the other structures and is visually similar to the highest dissipation structure seen at $Re=40$ in figure \ref{fig:burst_Re40_POs}, though whether these are states from the same solution branch has not been explored. 

\subsection{Condensate UPO at $Re=400$}
\begin{figure}
    \centering
    \includegraphics[width=\textwidth]{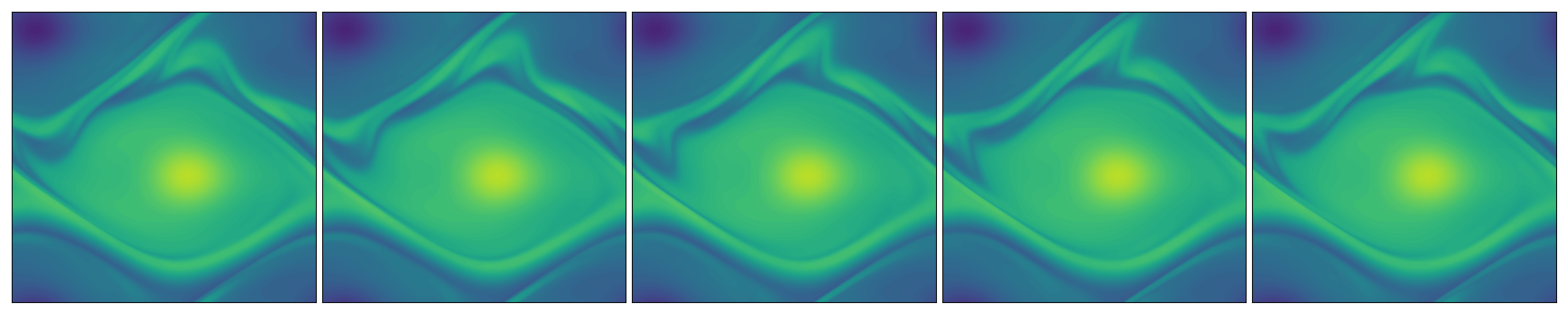}
    \caption{Snapshots of vorticity spaced every $T/6$ for the $l=1$ UPO found at $Re=400$. 
    The period is $T=0.4826$, and there is a small shift $s=0.0174$. 
    Vorticity contours in the figure run between $\pm 15$.}
    \label{fig:condensate_upos}
\end{figure}

At the highest Reynolds number, $Re=400$, even the high dissipation events are dominated by $l=1$ patterns (see figure \ref{fig:latent_burst}), consistent with the flow spending much of its time in a state dominated by a pair of opposite signed, large-scale vortices. 
We therefore do not attempt to find small scale UPOs here, but instead take random snapshots and project onto $l=1$ (as described in equation \ref{eqn:po_burst_guess}) as initial guesses in the Newton solver.
As expected, the success rate of this approach is very low, though we do find one UPO many times. 
This structure is shown in figure \ref{fig:condensate_upos} and consists of a large, near-stationary pair of vortices with an undulating vortex sheet between them. This structure is reminiscent of the `unimodal' state  discussed in \cite{Okamoto2010} who speculated that regardless of the forcing, such a universal solution always exists at high enough $Re$ in 2D. It is also, of course, consistent with the inverse cascade theory of 2D turbulence \citep[e.g.][]{Kraichnan1980}. The universal nature of the flow state suggests that it is converging to an Euler solution as $Re \rightarrow \infty$. This would mean, for example, that the single-vortex-pair state becomes unpinned  to the discrete translational symmetry in $y$ of the forcing function and so could drift around in both directions i.e. there could be travelling wave states. This type of solution has been seen recently in \cite{Zhigunov2023} albeit at much higher $Re$ ($=10^5$) and a smaller-wavelength forcing which oscillates in both spatial directions.
% \\

% \emph{TODO stability calculation -- section may be dropped/merged/moved to appendix. \rk{I wouldn't bother - lets see if a referee asks for this.}}

% Some early success with $l=1$ projections at the highest $Re$ (one solution so far). 
% Section will either be a full section if something significant, or a final few sentences on the previous section noting we find one or two solutions here.

%
% CONCLUSION
%
\section{Conclusion}
\label{sec:conc}
In this study we have built deep convolutional autoencoders to construct low-dimensional representations of Kolmogorov flow for a range of Reynolds numbers, $40 \leq Re \leq 400$.
Our architecture and training protocol were motivated by both (i) a discrete symmetry in the flow and (ii) a desire to maintain accuracy in the embedding even at high dissipation values. 
The resulting models were able to accurately reconstruct the original vorticity fields over all dissipation rates, even at the highest $Re=400$. %and we were able to use the internal latent representations in our networks to 
We then applied latent Fourier analysis to our trained networks to demonstrate that the learned embeddings are based on a sparse set of patterns associated with a small number of latent wavenumbers. 

% P2 changing latent structure, bursting events and the role of large-scale structure.
The latent representations of the turbulence were examined with dimensionality reduction techniques to reveal a single class of high dissipation event at $Re=40$, which was detached from the low-dissipation dynamics. 
The $Re=40$ bursts coincide with a much-weakened projection onto large scale structure and the emergence of small scale vortices associated with latent wavenumbers $l=2$ and $l=3$. 
At $Re=100$, the same approach revealed that the high- and low-dissipation dynamics are no longer distinct and can be smoothly interpolated between in the latent space. 
High-dissipation bursts, while still exciting patterns associated with $l=2$ and 3, also feature larger-scale structure with a lengthscale on the order of the domain size. 
Finally, the highest $Re=400$ is completely dominated by a domain filling vortex pair, even during high dissipation bursts. 

% P3 simple invariant solutions -- spatial localisation and future work. Implications for multiscale structure, and also the condensate PO. 
We then used the latent representations of turbulent trajectories to generate guesses for UPOs, initially by improving on a traditional recurrent flow analysis as $Re=40$, but also by using individual latent wavenumbers to generate guesses for high-dissipation solutions. 
We found very large numbers of high-dissipation periodic orbits at $Re=40$ using both $l=2$ and $l=3$ patterns as starting guesses. 
The former choice produced solutions which appear to be close to high dissipation events in the full turbulence, the latter produced UPOs which are clearly off-attractor. 
This suggests a multiscale picture where the triply-periodic $l=3$ structures are observed \emph{locally} alongside other larger-scale dynamics. 
This was also found with the high-dissipation, doubly-periodic UPOs converged at $Re=100$, and is consistent with the bursting events there showing evidence of large-scale structure ($l=1$) in the latent projections. 

The results emphasise the utility of low-order representations learned using large neural networks in understanding complex spatiotemporal dynamics. 
A decomposition into latent Fourier modes is not only a useful interpretability tool for exactly what the learnt basis represents, but is also a powerful technique for isolating small-scale exact solutions which capture dynamical events observed locally in the full turbulence. 
% The challenge now is 
With this picture, the challenge now is to find solutions which are themselves multiscale, or to understand the `rules' by which smaller-scale UPOs can coexist alongside larger-scale events. 
% The application of latent Fourier analysis to a three-dimensional flow as $Re$ increases may also be quite insightful, b 

\vspace{0.5cm}
\noindent
\textbf{Declaration of Interests.} The authors report no conflict of interest.

\vspace{0.5cm}
\noindent
\textbf{Acknowledgements.} JP acknowledges support from a UKRI Frontier Guarantee Grant EP/Y004094/1. 
JH acknowledges the support of Cantab Capital Institute for the Mathematics of Information.
MPB acknowledges support from the Office of Naval Research (ONR N00014-17-1-3029) and the NSF AI Institute of Dynamic Systems (\#2112085).

\appendix
\section{Network architecture and training}
\label{sec:app_network}
The changes in architecture relative to the simple feed-forward networks constructed in \citet{Page2021} can be summarised as follows:
\begin{enumerate}
    \item Replacing single convolutional layers with ``dense blocks'' \citep{huang2017densely}
    \item Placing batch normalisation layers between convolutions \citep{batchnorm}
    \item Using pure convolutions throughout without switching to fully connected blocks
    \item Using `GELU' activation functions throughout the network \citep{gelu_arxiv}, apart from the final (output) layer where $\tanh$ is preferred 
\end{enumerate}

Dense blocks \citep{huang2017densely,huang2019convolutional} are groups of convolutional layers where the output of each convolution operation is concatenated with its input, so the number of feature maps after the convolution and concatenation is the sum of the feature maps in the input upstream and those of the convolutional layer.
So if the input to the convolution is an `image' $\mathbf u$, which has shape $(N_x, N_y, N_{c1})$, where $N_{c1}$ is the number of channels, and the convolution operation produces an `image' $\mathbf v$ with shape $(N_x, N_y, N_{c2})$, then after concatenation we have an image of shape $(N_x, N_y, N_{c1} + N_{c2})$. 
We apply `periodic' padding to our images so that the discrete convolutions produce output with the same shape as the input. 
Our dense blocks always consist of three convolutional layers, where each convolutional layer adds 32 channels to the output. So if the input to the dense block has shape $(N_x, N_y, N_c)$, the output has shape $(N_x, N_y, N_c + 3 \times 32)$. 
After the dense block we apply another convolutional operation to reduce the number of channels, typically to 32 (exact architecture is summarised below). 

The `GELU' activation function \citep{gelu_arxiv} was designed to overcome some known problems with the more widely used `ReLU' activation function, particularly the occurrence of dead neurons within an architecture which are common in deep networks. 
This is done by removing the hard zero for negative inputs to the activation. The GELU activation is defined as
\begin{equation}
    \text{GELU}(x) := x \Phi(x),
\end{equation}
where $\Phi(x) = \mathbb P(X \leq x)$, with $X \sim N(0,1)$ is the cumulative distribution function for the standard normal distribution. 
The alternative activation $\tanh(x)$ is used at the final output layer for a symmetric output that can match the normalised input to the network $\omega(x,y) / \omega_{\text{norm}} \in [-1, 1]$.

Our encoding architecture then consists of a repeated sequence of (i) convolution, (ii) dense block, (iii) max pooling, until the final encoding layer where an additional convolution is used to produce an `image' of shape $(4, 8, M)$, where $1 \leq M \leq 32$ is a specified number of feature maps. 
The final streamwise dimension (4) is selected in an attempt to minimize the maximum latent wavenumber required (i.e. to compress the input into large-scale patterns), while we retain 8 cells in the vertical to match the shift-reflect symmetry in the system. 

Overall, the network operations in the encoder can be summarised as follows:
\begin{align*}
    \omega \to &\text{PC}(8\times 8, 64) \to \text{DB}(8 \times 8) \to \text{MP}(2,2) \to \\
    &\text{PC}(4\times 4, 32) \to \text{DB}(4 \times 4) \to \text{MP}(2,2) \to \\
    &\text{PC}(4\times 4, 32) \to \text{DB}(4 \times 4) \to \text{MP}(2,2) \to \\
    &\text{PC}(2\times 2, 32) \to \text{DB}(2 \times 2) \to \text{MP}(2,2) \to \\
    &\text{PC}(2\times 2, 32) \to \text{DB}(2 \times 2) \to \text{MP}(2,1) \to \\
    &\text{PC}(2\times 2, 32) \to \text{DB}(2\times 2) \to \text{PC}(2\times 2, M ) \equiv \mathscr E(\omega),
\end{align*}
where the terms in brackets represent the size of the convolutional filters, followed by the number of feature maps, `PC' standards for a `periodic convolution' (periodic padding on the image), `DB' for the dense block described above and `MP' stands for `max pooling' with the cell size given in brackets.
The input vorticity is a single-channel image of size $128\times 128$. 
The structure of the decoder is essentially the encoder described above, reversed (upsampling layers replace max pooling, and the final output layer has a single feature map). 

We tried many iterations of the architecture, including using fully connected layers near the encoding $\mathscr E$ and using pure feed forward networks.
Generally dropping either the pure fully convolutional aspects or removing the dense blocks reduced performance, with an increase of roughly an order of magnitude in the loss function.

When training we found the results to be highly sensitive to the learning rate in the Adam optimizer \citep{Kingma2015}. 
Values of $\eta \in \{10^{-5}, 10^{-4}, 3\times 10^{-4}, 5\times 10^{-4}, 10^{-3} \}$ were tried and $\eta = 5\times 10^{-4}$ was universally the best performing choice across all values of $Re$ and encoding dimensions. 

We trained each network for 500 epochs, training 3 identical architectures for most $Re$ and $m$ and selecting the best performing (in terms of training loss). 
A batch size of 64 was used throughout.
Overfitting was observed in around $1/3$ of cases, and we used early stopping based on the validation loss to extract the `best' weights from within the training process.

% \section{Details of simple invariant solutions}
% Summarise structures. Stability? If time -- only interested in condensate really. 

\bibliographystyle{jfm}

\begin{thebibliography}{57}
\expandafter\ifx\csname natexlab\endcsname\relax\def\natexlab#1{#1}\fi
\def\au#1{#1} \def\ed#1{#1} \def\yr#1{#1}\def\at#1{#1}\def\jt#1{\textit{#1}}
  \def\bt#1{#1}\def\bvol#1{\textbf{#1}} \def\vol#1{#1} \def\pg#1{#1}
  \def\publ#1{#1}\def\arxiv#1{#1}\def\org#1{#1}\def\st#1{\textit{#1}}

\bibitem[Artuso {\em et~al.\/}(1990{\natexlab{{\em a\/}}})Artuso, Aurell \&
  Cvitanovic]{Artuso1990a}
{\sc \au{Artuso, R.}, \au{Aurell, E.} \& \au{Cvitanovic, P.}}
  \yr{1990{\natexlab{{\em a\/}}}}  \at{{Recycling of strange sets: I cycle
  expansions}}.  \jt{Nonlinearity}  \bvol{3},  \pg{325--359}.

\bibitem[Artuso {\em et~al.\/}(1990{\natexlab{{\em b\/}}})Artuso, Aurell \&
  Cvitanovic]{Artuso1990b}
{\sc \au{Artuso, R.}, \au{Aurell, E.} \& \au{Cvitanovic, P.}}
  \yr{1990{\natexlab{{\em b\/}}}}  \at{{Recycling of strange sets: II
  applications}}.  \jt{Nonlinearity}  \bvol{3},  \pg{361--386}.

\bibitem[Boffetta \& Ecke(2012)]{Boffetta2012}
{\sc \au{Boffetta, Guido} \& \au{Ecke, Robert~E.}} \yr{2012}
  \at{Two-dimensional turbulence}.  \jt{Annual Review of Fluid Mechanics}
  \bvol{44}~(1),  \pg{427--451}.

\bibitem[Brunton {\em et~al.\/}(2020)Brunton, Noack \&
  Koumoutsakos]{Brunton2020}
{\sc \au{Brunton, Steven~L.}, \au{Noack, Bernd~R.} \& \au{Koumoutsakos,
  Petros}} \yr{2020}  \at{Machine learning for fluid mechanics}.  \jt{Annual
  Review of Fluid Mechanics}  \bvol{52}~(1),  \pg{477--508}.

\bibitem[Chandler \& Kerswell(2013)]{Chandler2013}
{\sc \au{Chandler, G.~J.} \& \au{Kerswell, R.~R.}} \yr{2013}  \at{Invariant
  recurrent solutions embedded in a turbulent two-dimensional {K}olmogorov
  flow}.  \jt{Journal of Fluid Mechanics}  \bvol{722},  \pg{554–595}.

\bibitem[Crowley {\em et~al.\/}(2022)Crowley, Pughe-Sanford, Toler, Krygier,
  Grigoriev \& Schatz]{Crowley2022}
{\sc \au{Crowley, C.~J.}, \au{Pughe-Sanford, J.~L.}, \au{Toler, W.},
  \au{Krygier, M.~C.}, \au{Grigoriev, R.~O.} \& \au{Schatz, M.~F.}} \yr{2022}
  \at{Turbulence tracks recurrent solutions}.  \jt{Proc. Nat. Acad. Sci.}
  \bvol{119},  \pg{e2120665119}.

\bibitem[Cvitanovi{\'c} {\em et~al.\/}(2016)Cvitanovi{\'c}, Artuso, Mainieri,
  Tanner \& Vattay]{ChaosBook}
{\sc \au{Cvitanovi{\'c}, P.}, \au{Artuso, R.}, \au{Mainieri, R.}, \au{Tanner,
  G.} \& \au{Vattay, G.}} \yr{2016} {\em Chaos: Classical and Quantum\/}.
  \publ{Copenhagen: Niels Bohr Inst.}

\bibitem[Cvitanovic \& Gibson(2010)]{Cvitanovic2010}
{\sc \au{Cvitanovic, P.} \& \au{Gibson, J.~F.}} \yr{2010}  \at{{ Geometry of
  the turbulence in wall-bounded shear flows: periodic orbits }}.  \jt{Physica
  Scripta}  \bvol{T142},  \pg{014007}.

\bibitem[De~Jes\'us \& Graham(2023)]{JesusGraham2023}
{\sc \au{De~Jes\'us, Carlos E.~P\'erez} \& \au{Graham, Michael~D.}} \yr{2023}
  \at{Data-driven low-dimensional dynamic model of kolmogorov flow}.  \jt{Phys.
  Rev. Fluids}  \bvol{8},  \pg{044402}.

\bibitem[Doohan {\em et~al.\/}(2019)Doohan, Willis \& Hwang]{Doohan2019}
{\sc \au{Doohan, Patrick}, \au{Willis, Ashley~P.} \& \au{Hwang, Yongyun}}
  \yr{2019}  \at{Shear stress-driven flow: the state space of near-wall
  turbulence as}.  \jt{Journal of Fluid Mechanics}  \bvol{874},  \pg{606--638}.

\bibitem[Dresdner {\em et~al.\/}(2022)Dresdner, Kochkov, Norgaard,
  Zepeda-Núñez, Smith, Brenner \& Hoyer]{Dresdner2022}
{\sc \au{Dresdner, Gideon}, \au{Kochkov, Dmitrii}, \au{Norgaard, Peter},
  \au{Zepeda-Núñez, Leonardo}, \au{Smith, Jamie~A.}, \au{Brenner, Michael~P.}
  \& \au{Hoyer, Stephan}} \yr{2022}  \at{Learning to correct spectral methods
  for simulating turbulent flows} .

\bibitem[Eckhardt {\em et~al.\/}(2002)Eckhardt, Faisst, Schmiegel \&
  Schumacher]{Eckhardt2002}
{\sc \au{Eckhardt, B.}, \au{Faisst, H.}, \au{Schmiegel, A.} \& \au{Schumacher,
  J.}} \yr{2002}  \at{{Turbulence transition in shear flows}}.  \jt{Advances in
  Turbulence IX: Proceedings 9th European Turbulence Conference (Southampton)
  (ed. I. P. Castro et al.) CISME}  \bvol{1},  \pg{701}.

\bibitem[Eckhardt {\em et~al.\/}(2007)Eckhardt, Schneider, Hof \&
  Westerweel]{Eckhardt2007}
{\sc \au{Eckhardt, B.}, \au{Schneider, T.~M.}, \au{Hof, B.} \& \au{Westerweel,
  J.}} \yr{2007}  \at{{ Turbulence transition in pipe flow }}.  \jt{Annual
  Review of Fluid Mechanics}  \bvol{39},  \pg{447--468}.

\bibitem[Farazmand(2016)]{Farazmand2016}
{\sc \au{Farazmand, M.}} \yr{2016}  \at{An adjoint-based approach for finding
  invariant solutions of navier{\textendash}stokes equations}.  \jt{Journal of
  Fluid Mechanics}  \bvol{795},  \pg{278--312}.

\bibitem[Gibson \& Brand(2014)]{GibsonBrand2014}
{\sc \au{Gibson, J.~F.} \& \au{Brand, E.}} \yr{2014}  \at{{ Spanwise-localized
  solutions of planar shear flows}}.  \jt{Journal of Fluid Mechanics}
  \bvol{745},  \pg{25--61}.

\bibitem[Gibson {\em et~al.\/}(2008)Gibson, Halcrow \& Cvitanovic]{Gibson2008}
{\sc \au{Gibson, J.~F.}, \au{Halcrow, J.} \& \au{Cvitanovic, P.}} \yr{2008}
  \at{Visualizing the geometry of state space in plane couette flow}.
  \jt{Journal of Fluid Mechanics}  \bvol{611},  \pg{107–130}.

\bibitem[Graham \& Floryan(2021)]{Graham2021}
{\sc \au{Graham, Michael~D.} \& \au{Floryan, Daniel}} \yr{2021}  \at{Exact
  coherent states and the nonlinear dynamics of wall-bounded turbulent flows}.
  \jt{Annual Review of Fluid Mechanics}  \bvol{53}~(1),  \pg{227--253}.

\bibitem[{Gudorf} \& {Cvitanovic}(2019)]{Gudorf2019}
{\sc \au{{Gudorf}, Matthew} \& \au{{Cvitanovic}, Predrag}} \yr{2019}
  {Spatiotemporal tiling of the Kuramoto-Sivashinsky equation}.  \bt{In {\em
  APS March Meeting Abstracts\/}},  \st{APS Meeting Abstracts},  \vol{vol.
  2019},  \pg{p. L70.263}.

\bibitem[Hall \& Sherwin(2010)]{Hall2010}
{\sc \au{Hall, P.} \& \au{Sherwin, S.}} \yr{2010}  \at{Streamwise vortices in
  shear flows: harbingers of transition and the skeleton of coherent
  structures}.  \jt{Journal of Fluid Mechanics}  \bvol{661},  \pg{178–205}.

\bibitem[Hendrycks \& Gimpel(2016)]{gelu_arxiv}
{\sc \au{Hendrycks, Dan} \& \au{Gimpel, Kevin}} \yr{2016} Gaussian error linear
  units (gelus).

\bibitem[Hopf(1948)]{Hopf1948}
{\sc \au{Hopf, E.}} \yr{1948}  \at{A mathematical example displaying features
  of turbulence}.  \jt{Commun. Pure Appl. Math.}  \bvol{1},  \pg{303--322}.

\bibitem[Huang {\em et~al.\/}(2017)Huang, Liu, van~der Maaten \&
  Weinberger]{huang2017densely}
{\sc \au{Huang, Gao}, \au{Liu, Zhuang}, \au{van~der Maaten, Laurens} \&
  \au{Weinberger, Kilian~Q}} \yr{2017} Densely connected convolutional
  networks.  \bt{In {\em Proceedings of the IEEE Conference on Computer Vision
  and Pattern Recognition\/}}.

\bibitem[Huang {\em et~al.\/}(2019)Huang, Liu, Pleiss, Van Der~Maaten \&
  Weinberger]{huang2019convolutional}
{\sc \au{Huang, Gao}, \au{Liu, Zhuang}, \au{Pleiss, Geoff}, \au{Van Der~Maaten,
  Laurens} \& \au{Weinberger, Kilian}} \yr{2019}  \at{Convolutional networks
  with dense connectivity}.  \jt{IEEE Transactions on Pattern Analysis and
  Machine Intelligence} .

\bibitem[Ioffe \& Szegedy(2015)]{batchnorm}
{\sc \au{Ioffe, Sergey} \& \au{Szegedy, Christian}} \yr{2015} Batch
  normalization: Accelerating deep network training by reducing internal
  covariate shift.  \bt{In {\em Proceedings of the 32nd International
  Conference on Machine Learning\/} (ed. \ed{Francis Bach \& David Blei})},
  \st{Proceedings of Machine Learning Research},  \vol{vol.~37},  \pg{pp.
  448--456}.  \publ{Lille, France: PMLR}.

\bibitem[Jim{\'{e}}nez(2012)]{Jimenez2012}
{\sc \au{Jim{\'{e}}nez, Javier}} \yr{2012}  \at{Cascades in wall-bounded
  turbulence}.  \jt{Annual Review of Fluid Mechanics}  \bvol{44}~(1),
  \pg{27--45}.

\bibitem[Jim{\'{e}}nez(2020)]{Jimenez2020}
{\sc \au{Jim{\'{e}}nez, Javier}} \yr{2020}  \at{Dipoles and streams in
  two-dimensional turbulence}.  \jt{Journal of Fluid Mechanics}  \bvol{904}.

\bibitem[Kawahara \& Kida(2001)]{Kawahara2001}
{\sc \au{Kawahara, G.} \& \au{Kida, S.}} \yr{2001}  \at{Periodic motion
  embedded in plane couette turbulence: regeneration cycle and burst}.
  \jt{Journal of Fluid Mechanics}  \bvol{449},  \pg{291–300}.

\bibitem[Kawahara {\em et~al.\/}(2012)Kawahara, Uhlmann \& van
  Veen]{Kawahara2012}
{\sc \au{Kawahara, G.}, \au{Uhlmann, M.} \& \au{van Veen, L.}} \yr{2012}
  \at{The significance of simple invariant solutions in turbulent flows}.
  \jt{Annual Review of Fluid Mechanics}  \bvol{44}~(1),  \pg{203--225}.

\bibitem[Kerswell(2005)]{Kerswell2005}
{\sc \au{Kerswell, R.~R.}} \yr{2005}  \at{{ Recent progress in understanding
  the transition to turbulence in a pipe}}.  \jt{Nonlinearity}  \bvol{18},
  \pg{R17--R44}.

\bibitem[Kim \& Okamoto(2010)]{Okamoto2010}
{\sc \au{Kim, S.-C.} \& \au{Okamoto, H.}} \yr{2010}  \at{Vortices of large
  scale appearing in the 2d stationary navier-stokes equations at large
  reynolds numbers}.  \jt{Japan J. Indust. Appl. Math.}  \bvol{27},
  \pg{47--71}.

\bibitem[Kingma \& Ba(2015)]{Kingma2015}
{\sc \au{Kingma, Diederik~P.} \& \au{Ba, Jimmy}} \yr{2015} Adam: {A} method for
  stochastic optimization.  \bt{In {\em 3rd International Conference on
  Learning Representations, {ICLR} 2015, San Diego, CA, USA, May 7-9, 2015,
  Conference Track Proceedings\/} (ed. \ed{Yoshua Bengio \& Yann LeCun})}.

\bibitem[Kochkov {\em et~al.\/}(2021)Kochkov, Smith, Alieva, Wang, Brenner \&
  Hoyer]{Kochkov2021}
{\sc \au{Kochkov, D.}, \au{Smith, J.~A.}, \au{Alieva, A.}, \au{Wang, Q.},
  \au{Brenner, M.~P.} \& \au{Hoyer, S.}} \yr{2021}  \at{Machine
  learning-accelerated computational fluid dynamics}.  \jt{Proceedings of the
  National Academy of Sciences}  \bvol{118},  \pg{e2101784118}.

\bibitem[Kraichnan \& Montgomery(1980)]{Kraichnan1980}
{\sc \au{Kraichnan, R~H} \& \au{Montgomery, D}} \yr{1980}  \at{Two-dimensional
  turbulence}.  \jt{Reports on Progress in Physics}  \bvol{43}~(5),
  \pg{547--619}.

\bibitem[Lan \& Cvitanovic(2004)]{Lan2004}
{\sc \au{Lan, Y.~H.} \& \au{Cvitanovic, P.}} \yr{2004}  \at{Variational method
  for finding periodic orbits in a general flow}.  \jt{Phys. Rev. E}
  \bvol{69},  \pg{016217}.

\bibitem[LeCun {\em et~al.\/}(2015)LeCun, Bengio \& Hinton]{LeCun2015}
{\sc \au{LeCun, Yann}, \au{Bengio, Yoshua} \& \au{Hinton, Geoffrey}} \yr{2015}
  \at{Deep learning}.  \jt{Nature}  \bvol{521}~(7553),  \pg{436--444}.

\bibitem[Linot \& Graham(2020)]{Linot2020}
{\sc \au{Linot, Alec~J.} \& \au{Graham, Michael~D.}} \yr{2020}  \at{Deep
  learning to discover and predict dynamics on an inertial manifold}.
  \jt{Physical Review E}  \bvol{101},  \pg{062209}.

\bibitem[Linot \& Graham(2023)]{Linot2023}
{\sc \au{Linot, Alec~J.} \& \au{Graham, Michael~D.}} \yr{2023} Dynamics of a
  data-driven low-dimensional model of turbulent minimal couette flow.

\bibitem[Lucas \& Kerswell(2014)]{Lucas2014}
{\sc \au{Lucas, Dan} \& \au{Kerswell, Rich}} \yr{2014}  \at{Spatiotemporal
  dynamics in two-dimensional kolmogorov flow over large domains}.  \jt{Journal
  of Fluid Mechanics}  \bvol{750},  \pg{518–554}.

\bibitem[Lucas \& Kerswell(2015)]{Lucas2015}
{\sc \au{Lucas, D.} \& \au{Kerswell, R.~R.}} \yr{2015}  \at{Recurrent flow
  analysis in spatiotemporally chaotic 2-dimensional {K}olmogorov flow}.
  \jt{Physics of Fluids}  \bvol{27},  \pg{045106}.

\bibitem[van~der Maaten \& Hinton(2008)]{tsne}
{\sc \au{van~der Maaten, L. J.~P.} \& \au{Hinton, G.~E.}} \yr{2008}
  \at{Visualizing high-dimensional data using t-{S}{N}{E}}.  \jt{Journal of
  Machine Learning Research}  \bvol{9},  \pg{2579--2605}.

\bibitem[Marensi {\em et~al.\/}(2022)Marensi, Yaln{\i}z, Hof \&
  Budanur]{Marensi2022}
{\sc \au{Marensi, E.}, \au{Yaln{\i}z, G.}, \au{Hof, B.} \& \au{Budanur, N.B.}}
  \yr{2022}  \at{Symmetry-reduced dynamic mode decomposition of near-wall
  turbulence}.  \jt{Journal of Fluid Mechanics}  \bvol{954}.

\bibitem[{McInnes} {\em et~al.\/}(2018){McInnes}, {Healy} \&
  {Melville}]{2018arXivUMAP}
{\sc \au{{McInnes}, L.}, \au{{Healy}, J.} \& \au{{Melville}, J.}} \yr{2018}
  \at{{UMAP: Uniform Manifold Approximation and Projection for Dimension
  Reduction}}.  \jt{ArXiv e-prints} ,  \arxiv{arXiv: 1802.03426}.

\bibitem[Onsager(1949)]{Onsager1949}
{\sc \au{Onsager, L.}} \yr{1949}  \at{Statistical hydrodynamics}.  \jt{Il Nuovo
  Cimento}  \bvol{6}~(S2),  \pg{279--287}.

\bibitem[Page {\em et~al.\/}(2021)Page, Brenner \& Kerswell]{Page2021}
{\sc \au{Page, J.}, \au{Brenner, M.~P.} \& \au{Kerswell, R.~R.}} \yr{2021}
  \at{Revealing the state space of turbulence using machine learning}.
  \jt{Physical Review Fluids}  \bvol{6},  \pg{034402}.

\bibitem[Page \& Kerswell(2020)]{Page2020}
{\sc \au{Page, Jacob} \& \au{Kerswell, Rich~R.}} \yr{2020}  \at{Searching
  turbulence for periodic orbits with dynamic mode decomposition}.  \jt{Journal
  of Fluid Mechanics}  \bvol{886}.

\bibitem[Page {\em et~al.\/}(2022)Page, Norgaard, Brenner \&
  Kerswell]{PNBK2022}
{\sc \au{Page, Jacob}, \au{Norgaard, Peter}, \au{Brenner, Michael~P.} \&
  \au{Kerswell, Rich~R.}} \yr{2022} Recurrent flow patterns as a basis for
  turbulence: predicting statistics from structures.

\bibitem[Parker \& Schneider(2022)]{Parker2022}
{\sc \au{Parker, J.P.} \& \au{Schneider, T.M.}} \yr{2022}  \at{Variational
  methods for finding periodic orbits in the incompressible
  navier{\textendash}stokes equations}.  \jt{Journal of Fluid Mechanics}
  \bvol{941}.

\bibitem[Rowley {\em et~al.\/}(2009)Rowley, Mezi\'c, Bagheri, Schlatter \&
  Henningson]{Rowley2009}
{\sc \au{Rowley, C.~W.}, \au{Mezi\'c, I.}, \au{Bagheri, S.}, \au{Schlatter, P.}
  \& \au{Henningson, D.~S.}} \yr{2009}  \at{{Spectral analysis of nonlinear
  flows}}.  \jt{J. Fluid Mech.}  \bvol{641},  \pg{115--127}.

\bibitem[Schmid(2010)]{Schmid2010}
{\sc \au{Schmid, P.~J.}} \yr{2010}  \at{{Dynamic mode decomposition of
  numerical and experimental data}}.  \jt{J. Fluid Mech.}  \bvol{656},
  \pg{5--28}.

\bibitem[Smith \& Yakhot(1993)]{Smith1993}
{\sc \au{Smith, Leslie~M.} \& \au{Yakhot, Victor}} \yr{1993}  \at{Bose
  condensation and small-scale structure generation in a random force driven 2d
  turbulence}.  \jt{Physical Review Letters}  \bvol{71}~(3),  \pg{352--355}.

\bibitem[Suri {\em et~al.\/}(2020)Suri, Kageorge, Grigoriev \&
  Schatz]{Suri2020}
{\sc \au{Suri, B.}, \au{Kageorge, L.}, \au{Grigoriev, R.~O.} \& \au{Schatz,
  M.~F.}} \yr{2020}  \at{{Capturing turbulent dynamics and statistics in
  expriments with unstable periodic orbits}}.  \jt{Phys. Rev. Lett.}
  \bvol{125},  \pg{064501}.

\bibitem[van Veen {\em et~al.\/}(2019)van Veen, Vela-Mart{\'{\i}}n, Kawahara \&
  Yasuda]{Veen2019}
{\sc \au{van Veen, Lennaert}, \au{Vela-Mart{\'{\i}}n, Alberto}, \au{Kawahara,
  Genta} \& \au{Yasuda, Tatsuya}} \yr{2019}  \at{Periodic orbits in large eddy
  simulation of box turbulence}.  \jt{Fluid Dynamics Research}  \bvol{51}~(1),
  \pg{011411}.

\bibitem[Viswanath(2007)]{Viswanath2007}
{\sc \au{Viswanath, D.}} \yr{2007}  \at{Recurrent motions within plane
  {C}ouette turbulence}.  \jt{Journal of Fluid Mechanics}  \bvol{580},
  \pg{339–358}.

\bibitem[Wang {\em et~al.\/}(2007)Wang, Gibson \& Waleffe]{Wang2007}
{\sc \au{Wang, J.}, \au{Gibson, J.} \& \au{Waleffe, F.}} \yr{2007}  \at{Lower
  branch coherent states in shear flows: {t}ransition and control}.
  \jt{Physical Review Letters}  \bvol{98},  \pg{204501}.

\bibitem[Willis {\em et~al.\/}(2013)Willis, Cvitanovic \& Avila]{Willis2013}
{\sc \au{Willis, A.~P.}, \au{Cvitanovic, P.} \& \au{Avila, M.}} \yr{2013}
  \at{{Revealing the state space of turbulent pipe flow by symmetry
  reduction}}.  \jt{Journal of Fluid Mechanics}  \bvol{721},  \pg{514--540}.

\bibitem[Yaln{\i}z {\em et~al.\/}(2021)Yaln{\i}z, Hof \& Budanur]{Yalnuz2021}
{\sc \au{Yaln{\i}z, G\"{o}khan}, \au{Hof, Bj\"{o}rn} \& \au{Budanur,
  Nazmi~Burak}} \yr{2021}  \at{Coarse graining the state space of a turbulent
  flow using periodic orbits}.  \jt{Physical Review Letters}  \bvol{126}~(24).

\bibitem[Zhigunov \& Grigoriev(2023)]{Zhigunov2023}
{\sc \au{Zhigunov, Dmitriy} \& \au{Grigoriev, Roman~O.}} \yr{2023}  \at{Exact
  coherent structures in fully developed two-dimensional turbulence}.
  \jt{Journal of Fluid Mechanics}  \bvol{970},  \pg{A18}.

\end{thebibliography}

%%
%% TABLES
%%
%% If there are any tables, put them here.
%%

\end{document}